\documentclass[12pt]{article}
\usepackage{amssymb,amsmath}
\usepackage{graphicx}
\usepackage{hyperref}
 \newcommand{\be}{\begin{eqnarray} } 
  \newcommand{\ee}{\end{eqnarray} } 
\newcommand {\unit} [1] {\; \mathrm {#1}}

\begin{document}
\baselineskip 0.6cm

\def\simgt{\mathrel{\lower2.5pt\vbox{\lineskip=0pt\baselineskip=0pt
           \hbox{$>$}\hbox{$\sim$}}}}
\def\simlt{\mathrel{\lower2.5pt\vbox{\lineskip=0pt\baselineskip=0pt
           \hbox{$<$}\hbox{$\sim$}}}}

\begin{titlepage}

\begin{flushright}
UCB-PTH-11/11 \\
\end{flushright}

\vskip 1.7cm

\begin{center}

{\Large \bf 
A Natural SUSY Higgs Near 125 GeV
}

\vskip 0.8cm

{\large Lawrence J. Hall, David Pinner and Joshua T. Ruderman}

\vskip 0.4cm

{\it Berkeley Center for Theoretical Physics, Department of Physics, \\
     and Theoretical Physics Group, Lawrence Berkeley National Laboratory, \\
     University of California, Berkeley, CA 94720, USA} \\

\abstract{
The naturalness of a Higgs boson with a mass near 125~GeV is explored in a variety of weak-scale supersymmetric models.  A Higgs mass of this size strongly points towards a non-minimal implementation of supersymmetry. 
The Minimal Supersymmetric Standard Model now requires large $A$-terms to avoid multi-TeV stops.  The fine-tuning is at least 1\% for low messenger scales, and an order of magnitude worse for high messenger scales.   Naturalness is significantly improved in theories with a singlet superfield $S$ coupled to the Higgs superfields via $\lambda S H_u H_d$.  If $\lambda$ is perturbative up to unified scales, a fine-tuning of about 10\% is possible with a low mediation scale.  Larger values of $\lambda$, implying new strong interactions below unified scales, allow for a highly natural 125~GeV Higgs boson over a wide range of parameters.  Even for $\lambda$ as large as 2, where a heavier Higgs might be expected, a light Higgs boson naturally results from singlet-doublet scalar mixing.  Although the Higgs is light, naturalness allows for stops as heavy as 1.5~TeV and a gluino as heavy as 3~TeV.   Non-decoupling effects among the Higgs doublets can significantly suppress the coupling of the light Higgs to $b$ quarks in theories with a large $\lambda$, enhancing the $\gamma \gamma$ and $WW$ signal rates at the LHC by an order one factor relative to the Standard Model Higgs.
}

\end{center}
\end{titlepage}

\section{Introduction}
\label{sec:intro}
The ATLAS and CMS Collaborations have recently presented the first evidence for a Higgs boson with a mass of 124--126~GeV \cite{AtlasTalk,CMSTalk}.  The $\gamma \gamma$ channel yields excesses at the 2--3~$\sigma$ level for ATLAS and CMS, insufficient for a clear discovery.  Yet the concordance between the ATLAS and CMS excesses increases the likelihood that this is indeed the Higgs boson, and motivates us to study the implications for natural electroweak breaking in the context of weak-scale supersymmetry.

In the Minimal Supersymmetric Standard Model (MSSM) the lightest Higgs boson is lighter than about 135 GeV, depending on top squark parameters (for a review with original references, see~\cite{Djouadi:2005gj}), and heavier than 114 GeV, the LEP bound on the Standard Model Higgs~\cite{Barate:2003sz}.  A Higgs mass of 125 GeV naively seems perfect, lying midway between the experimental lower bound and the theoretical upper limit.
The key motivation for weak-scale supersymmetry is the naturalness problem of the weak scale and therefore we take the degree of fine-tuning~\cite{CERN-TH-4825/87, hep-ph/0007265, SusyFineTuning, hep-ph/0602096, GoldenMaxMix}  as a crucial tool in guiding us to the most likely implementation of a 125 GeV Higgs.   
In this regard we find that increasing the Higgs mass from its present bound to 125 GeV has highly significant consequences.  In the limit of decoupling one Higgs doublet the light Higgs mass is given by
\begin{equation}
m_h^2 \, = \, M_Z^2 \cos^2 2 \beta + \delta_t^2
\label{eq:hmassMSSM}
\end{equation}
where $\delta_t^2$ arises from loops of heavy top quarks and top squarks and $\tan \beta$ is the ratio of electroweak vacuum expectation values.  At large $\tan \beta$, we require $\delta_t \approx 85$~GeV which means that a very substantial loop contribution, nearly as large as the tree-level mass, is required to raise the Higgs mass to 125 GeV.

\begin{figure}[h!]
\begin{center} \includegraphics[width=0.7\textwidth]{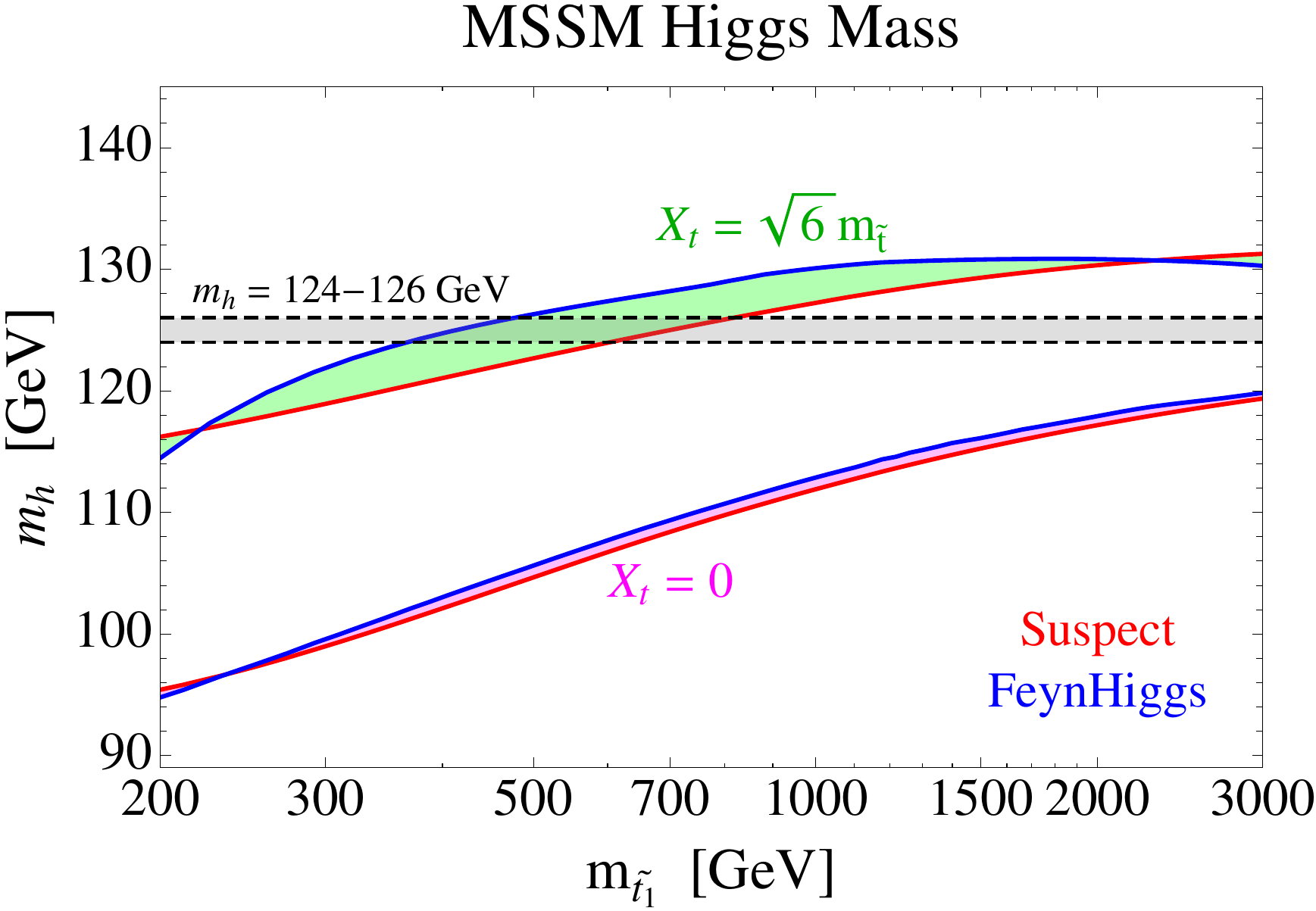} \end{center}
\caption{\label{fig:mssm-intro}
The Higgs mass in the MSSM as a function of the lightest top squark mass,  $m_{\tilde t_1}$, with red/blue solid lines computed using Suspect/FeynHiggs.  The two upper lines are for maximal top squark mixing assuming degenerate stop soft masses and yield a 124 (126) GeV Higgs mass for $m_{\tilde t_1}$ in the range of 350--600 (500--800) GeV, while the two lower lines are for zero top squark mixing and do not yield a 124 GeV Higgs mass for $m_{\tilde t_1}$ below 3 TeV.  Here we have taken $\tan\beta = 20$.  The shaded regions highlight the difference between the Suspect and FeynHiggs results, and may be taken as an estimate of the uncertainties in the two-loop calculation.
}
\end{figure}

The Higgs mass calculated at two loops in the MSSM is shown in Figure \ref{fig:mssm-intro} as a function of the lightest top squark mass for two values of the top squark mixing parameter $X_t$.  The red/blue contours are computed using the Suspect~\cite{suspect} and FeynHiggs~\cite{feynhiggs} packages, which have differing renormalization prescriptions and the spread between them, highlighted by the shading, may be taken as a rough measure of the current uncertainty in the calculation.  For a given Higgs mass, such as 125 GeV, large top squark mixing leads to lower and more natural top squark masses, although the mixing itself contributes to the fine-tuning, as we will discuss. In fact, stop mixing is required to raise the Higgs mass to 125 GeV without multi-TeV stops. Even at maximal mixing, we must have $\sqrt{m_{Q_3} m_{u_3}} \gtrsim 600$ GeV (which, for degenerate soft masses, results in stop masses heavier than have been directly probed by existing LHC searches~\cite{NaturalLHC,OtherLHCStop}) and, as we will discuss in the next section, this implies that fine-tuning of {\it at least 1\%} is required in the MSSM, even  for the extreme case of an ultra-low messenger scale of 10 TeV.  Hence we seek an alternative, more natural setting for a 125~GeV Higgs.

\begin{figure}[h!]
\begin{center} \includegraphics[width=0.7\textwidth]{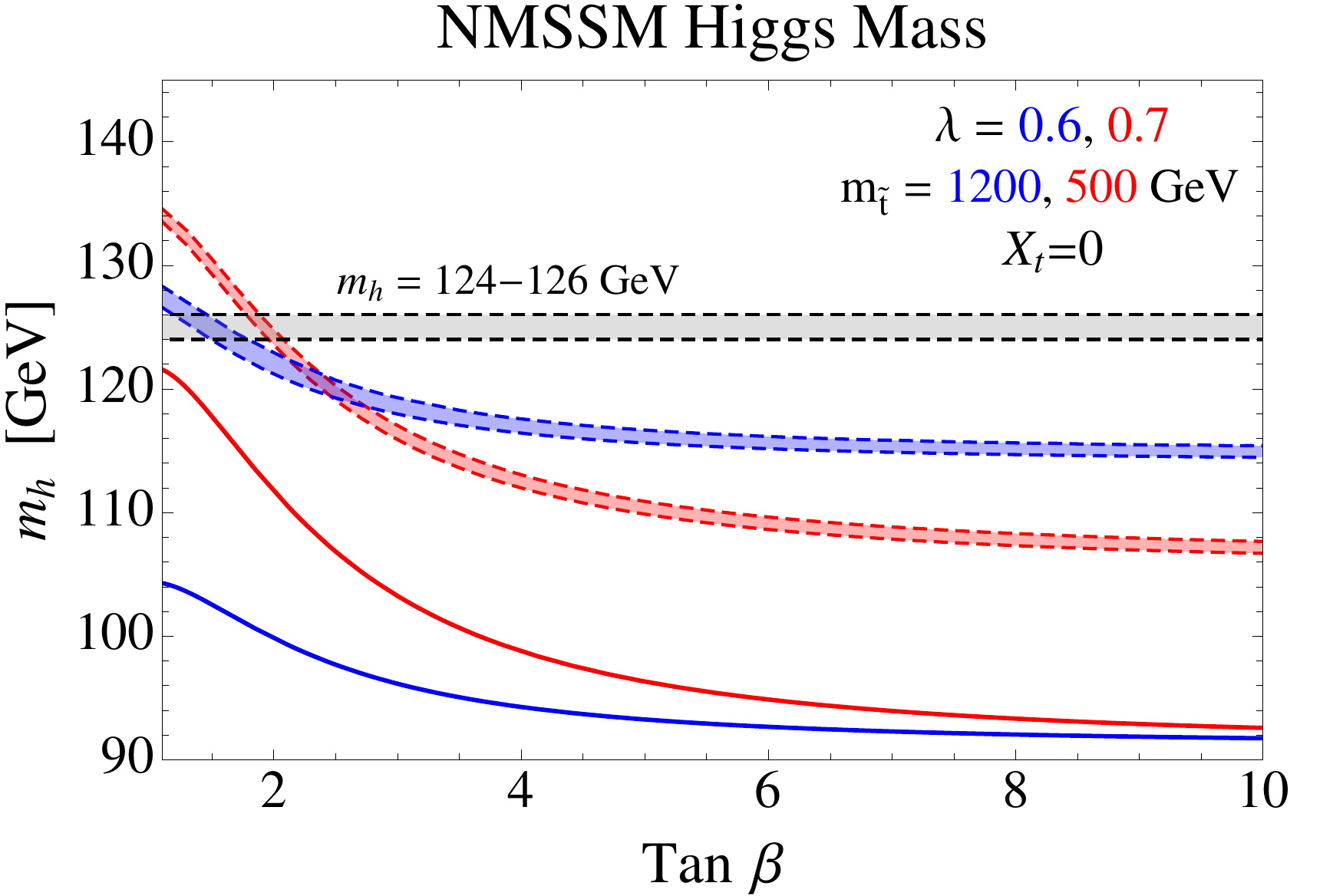} \end{center}
\caption{\label{fig:NMSSM_tbi}
The Higgs mass in the NMSSM as a function of $\tan \beta$.  The solid lines show the tree-level result of equation \ref{eq:hmassNMSSM} while the shaded bands bounded by dashed lines result from adding the $\lambda^2 v^2 \sin^2 2 \beta$ contribution of equation~\ref{eq:hmassNMSSM} to the two-loop Suspect/FeynHiggs MSSM result, with degenerate stop soft masses and no stop mixing.  The top contribution $\delta_t$ is sufficient to raise the Higgs mass to 125 GeV for $\lambda = 0.7$ for a top squark mass of 500 GeV; but as $\lambda$ is decreased to 0.6 a larger value of the top squark mass is needed.
}
\end{figure}

In the next-to-minimal model (NMSSM, for a review with references, see~\cite{Ellwanger:2009dp}) the supersymmetric Higgs mass parameter $\mu$ is promoted to a gauge-singlet superfield, $S$, with a coupling to the Higgs doublets, $\lambda S H_u H_d$, that is perturbative to unified scales, thereby constraining $\lambda \lesssim 0.7$ (everywhere in this paper $\lambda$ refers to the weak scale value of the coupling).  
The maximum mass of the lightest Higgs boson is
\begin{equation}
m_h^2 \, = \,  M_Z^2 \cos^2 2 \beta + \lambda^2 v^2 \sin^2 2 \beta + \delta_t^2,
\label{eq:hmassNMSSM}
\end{equation}
where here and throughout the paper we use $v=174$~GeV.  For $\lambda v > M_Z$, the tree-level contributions to $m_h$ are maximized for $\tan \beta=1$, as shown by the solid lines in Figure \ref{fig:NMSSM_tbi}, rather than by large values of $\tan \beta$ as in the MSSM.  However, even for $\lambda$ taking its maximal value of 0.7, these tree-level contributions cannot raise the Higgs mass above 122 GeV, and $\delta_t \gtrsim 28$~GeV is required.  Adding the top loop contributions allows the Higgs mass to reach 125~GeV, as shown by the shaded bands of Figure \ref{fig:NMSSM_tbi}, at least for low values of $\tan \beta$ in the region of 1--2.  In this case, unlike the MSSM, maximal stop mixing is not required to get the Higgs heavy enough.  In section 3 we demonstrate that, for a 125 GeV Higgs mass, the fine-tuning of the NMSSM is significantly improved relative to the MSSM, but only for $.6 \lesssim \lambda \lesssim .7$, near the boundary of perturbativity at the GUT scale.

In the ``$\lambda$-SUSY" theory \cite{hep-ph/0607332}, $\lambda$ is increased so that the interaction becomes non-perturbative below unified scales; but $\lambda$ should not exceed about 2, otherwise the non-perturbative physics occurs below 10 TeV and is likely to destroy the successful understanding of precision electroweak data in the perturbative theory.  The non-perturbativity of $\lambda$ notwithstanding, gauge coupling unification can be preserved in certain UV completions of $\lambda$-SUSY, such as the Fat Higgs~\cite{fathiggs}.  The $\lambda$-SUSY theory is highly motivated by an improvement in fine-tuning over the MSSM by roughly a factor of $2 \lambda^2/g^2 \sim 4 \lambda^2$, where $g$ is the SU(2) gauge coupling.  Equivalently, for the MSSM and $\lambda$-SUSY to have comparable levels of fine-tuning, the superpartner spectrum can be heavier in $\lambda$-SUSY by about a factor $2 \lambda$.  The origin of this improvement, a large value of $\lambda$ in the potential,  is correlated with the mass of the Higgs, which is naively raised from $gv/\sqrt{2}$ to $\lambda v$. However, this now appears to be excluded by current limits~\cite{CombinedHiggsHCP}, with $\lambda > 1$ giving a Higgs boson much heavier than 125~GeV (for other theories that raise the Higgs mass above that of the MSSM see~\cite{NonDecoupleDterm, hep-ph/0509244, arXiv:0707.0005}).

\begin{figure}[h!]
\begin{center} \includegraphics[width=0.7\textwidth]{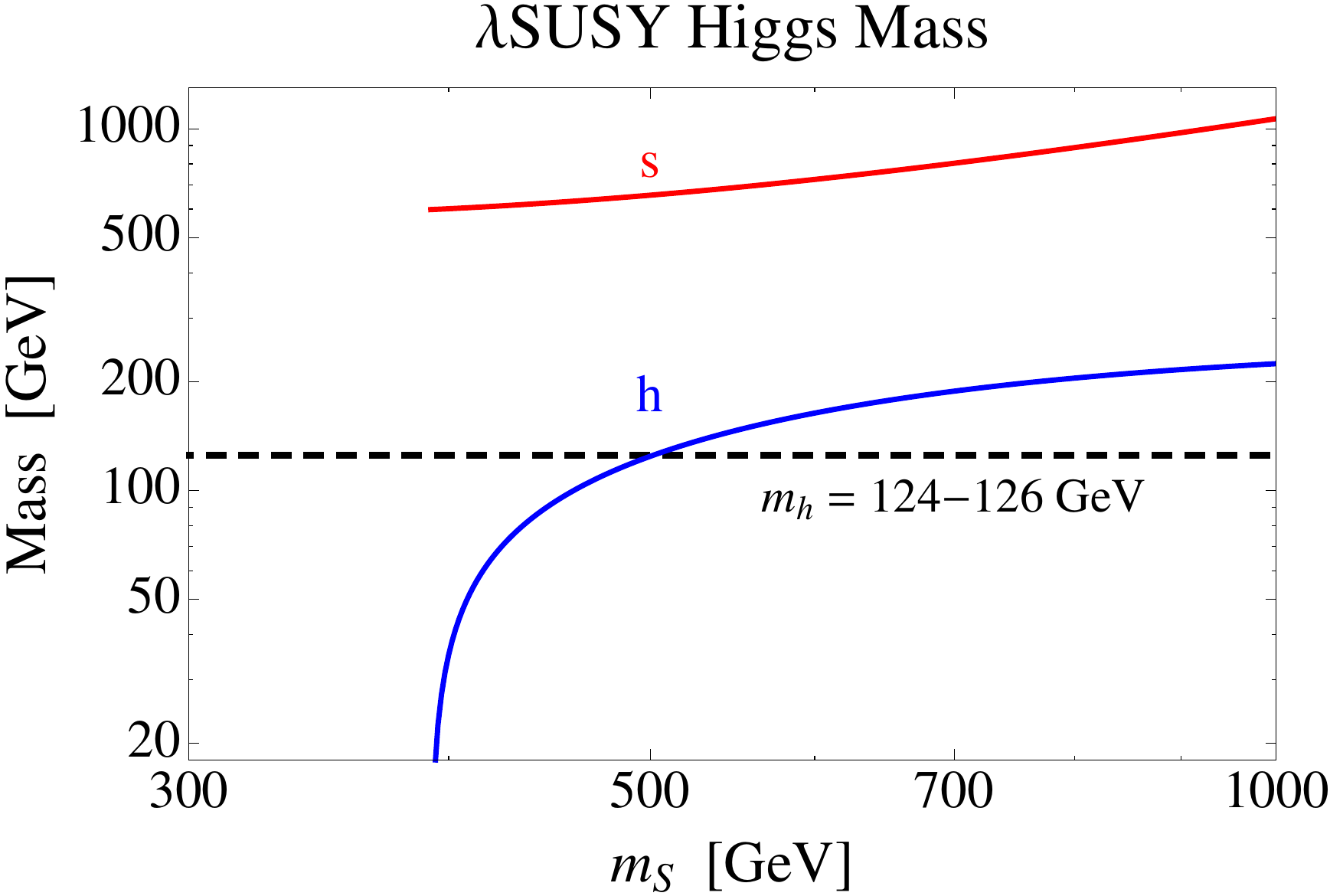} \end{center}
\caption{\label{fig:LambdaSUSYHiggs}
The Higgs mass in $\lambda$-SUSY, as a function of the singlet soft mass $m_S$.  Here, $\lambda = 2$, $\tan \beta =2$, and the other parameters are as described in Table~\ref{tab:bench}, which gives the light Higgs a mass of $m_h = 280$~GeV in the limit of heavy singlet mass.  However, we see that lowering the singlet mass $m_S$ results in a lighter Higgs due to mixing of the singlet with the Higgs.
}
\end{figure}

Most studies of $\lambda$-SUSY~\cite{hep-ph/0607332,LambdaSusyPheno} have decoupled the CP even singlet scalar $s$ by making its soft mass parameter, $m_S^2$, large.  This was often done purely for simplicity to avoid the complications of a $3 \times 3$ mass matrix for the CP even Higgs scalars.  However, this decoupling is itself unnatural since the soft Higgs doublet mass parameter is generated by one-loop renormalization group scaling at order $\lambda^2 m_S^2$.   For $\lambda = 2$, avoiding additional tuning at the 20\% level requires $m_S~\lesssim~1$~TeV~\cite{hep-ph/0607332}.  Once $s$ is no longer decoupled, it is crucial to include doublet-singlet Higgs mixing.  In the limit of decoupling one Higgs doublet, $s$ mixes with the remaining light neutral doublet Higgs $h$ at tree-level via the mass matrix
\begin{equation}
{\cal M}^2 \, = \, \left(
 \begin{matrix}
  \lambda^2 v^2 \sin^2 2 \beta + M_Z^2 \cos^2 2 \beta & \lambda v (\mu, M_S, A_\lambda) \\
   \lambda v (\mu, M_S, A_\lambda) & m_S^2 \end{matrix}
\right).
\label{eq:h-smassmixing}
\end{equation}
In general there are several contributions to the off-diagonal entry and these will be discussed in section \ref{sec:lambdasusy}; but all are proportional to $\lambda v$, which is large in $\lambda$-SUSY, so that mixing cannot be neglected even for rather large values of $m_S^2$.   This is illustrated in Figure \ref{fig:LambdaSUSYHiggs} where, for a set of reference parameters of the model discussed later, the two eigenvalues of this mixing matrix are shown as a function of $m_S$.  At the reference point $\lambda = 2$ and $\tan \beta = 2$, so that in the absence of mixing the Higgs mass would be 280 GeV, but this is reduced to 125 GeV for $m_S \sim$ 500 GeV.  As the blue curve of Figure~\ref{fig:LambdaSUSYHiggs} crosses 125~GeV its slope is quite modest -- a central claim of this paper is that a 125 GeV Higgs from doublet-singlet mixing in $\lambda$-SUSY is highly natural.  However, moving along the blue curve of Figure~\ref{fig:LambdaSUSYHiggs}, the tuning rapidly increases as the Higgs mass becomes lighter than 100~GeV.

The theory with $\lambda \sim 2$ and a light Higgs, due to singlet-doublet mixing, has a number of interesting consequences for LHC physics.  First of all, despite the light Higgs mass of 125~GeV, as discussed above the large value of $\lambda$ implies that the Higgs potential is less sensitive to corrections to the doublet Higgs soft masses, and the superpartners can be a factor of $2 \lambda \sim 4$ times heavier than in the MSSM before they spoil naturalness.  This means that stops can be as heavy as 1.5~TeV, and the gluino as heavy as 3~TeV, before fine-tuning reaches the 10\% level.  The usual interpretation~\cite{NaturalLHC} of the null results for supersymmetry at the LHC is that the stops should be lighter than the other squarks to maintain naturalness.  The situation in $\lambda$-SUSY is drastically different and we should not be surprised that we have not yet seen supersymmetry: the entire colored spectrum may be sitting, naturally, well above 1~TeV with flavor degenerate squarks!

$\lambda$-SUSY with a light Higgs boson also presents the possibility of interesting deformations of the SM Higgs branching ratios.  In $\lambda$-SUSY, we find that mixing between the light and heavy Higgs doublets leads to a depletion of the light Higgs coupling to bottom quarks, which has the effect of increasing the Higgs branching ratio to $\gamma \gamma$ and $WW$.  This is the opposite of the usual non-decoupling effect in the MSSM~\cite{Carena:2011fc}, where the Higgs coupling to bottoms is enhanced as the heavy Higgs mass decreases.  The effect is also numerically larger in $\lambda$-SUSY because the non-decoupling is enhanced by large $\lambda$.  We will see that, depending on parameters, the $gg \rightarrow h \rightarrow \gamma \gamma$ rate can be enhanced by up to about 50\% relative to the SM rate, in the most natural regime of parameter space.  Meanwhile, the branching ratio to bottoms can be depleted by a similar factor.  Usually, the conception is that supersymmetry should be first discovered by discovering superparticle production, however in $\lambda$-SUSY the first discovery of supersymmetry may be through exotic Higgs branching ratios.

The rest of the paper is structured as follows.  In section~ \ref{sec:MSSM} we discuss the implications of a 125 GeV Higgs boson for the MSSM and conclude that a Higgs of this mass disfavors the MSSM, motivating study of an alternative model.  In section~\ref{sec:NMSSM} we consider the implications for the NMSSM, where the fine-tuning can be significantly reduced relative to the MSSM, although only at the edge of the $(\lambda, \tan \beta)$ parameter space.  Then in section~\ref{sec:lambdasusy} we consider $\lambda$-SUSY and show that a light, 125 GeV Higgs boson emerges naturally from theories with a large value for $\lambda$.  Section~\ref{sec:concl} contains our conclusions.

\section{A Higgs Mass near 125 GeV in the MSSM}
\label{sec:MSSM}

In this section we review the Higgs sector of the MSSM~\cite{Djouadi:2005gj} and assess the consequences of a Higgs boson mass of 125 GeV.  We determine which parts of the parameter space allow for a Higgs mass of 125 GeV and how much fine-tuning is required.

The Higgs sector of the MSSM depends, at tree-level, on the ratio of the vevs, $\tan \beta$, and on the pseudoscalar mass $m_A$, which determines the mixing between the two CP even scalars.  In this section, we focus on the decoupling limit, $m_A \gg m_Z$, where the lightest CP even Higgs is SM-like in its coupling and has the largest possible tree-level mass (away from the decoupling limit, mixing drives the lightest mass eigenstate lighter).  In the decoupling limit, the tree-level Higgs mass is given by $m_Z \cos 2 \beta$ and is maximized at high $\tan \beta$, but is always far below 125~GeV. 

At the one-loop level, stops contribute to the Higgs mass and three more parameters become important, the stop soft masses, $m_{Q_3}$ and $m_{u_3}$, and the stop mixing parameter $X_t = A_t - \mu \cot \beta$.  The dominant one-loop contribution to the Higgs mass depends on the geometric mean of the stop masses, $m_{\tilde t}^2 = m_{Q_3} m_{u_3}$, and is given by,
\be \label{eq:HiggsLoop}
m_h^2 \approx m_Z^2 \cos^2 2\beta + \frac{3}{(4 \pi)^2} \frac{m_t^4}{v^2} \left[ \ln \frac{m_{\tilde t}^2}{m_t^2} + \frac{X_t^2}{m_{\tilde t}^2} \left(1-\frac{X_t^2}{12 m_{\tilde t}^2} \right)\right].
\ee
The Higgs mass is sensitive to the degree of stop mixing through the second term in the brackets, and is maximized for $|X_t| = X_t^{\rm max} =  \sqrt 6 \, m_{\tilde t}$, which is referred to as ``maximal mixing."  The Higgs mass depends logarithmically on the stop masses, which means, of course, that the necessary stop mass depends exponentially on the Higgs mass.   Therefore, an accurate loop calculation is essential in order to determine which stop mass corresponds to a 125 GeV Higgs.  

\begin{figure}[h!]
\includegraphics[width=1.0\textwidth]{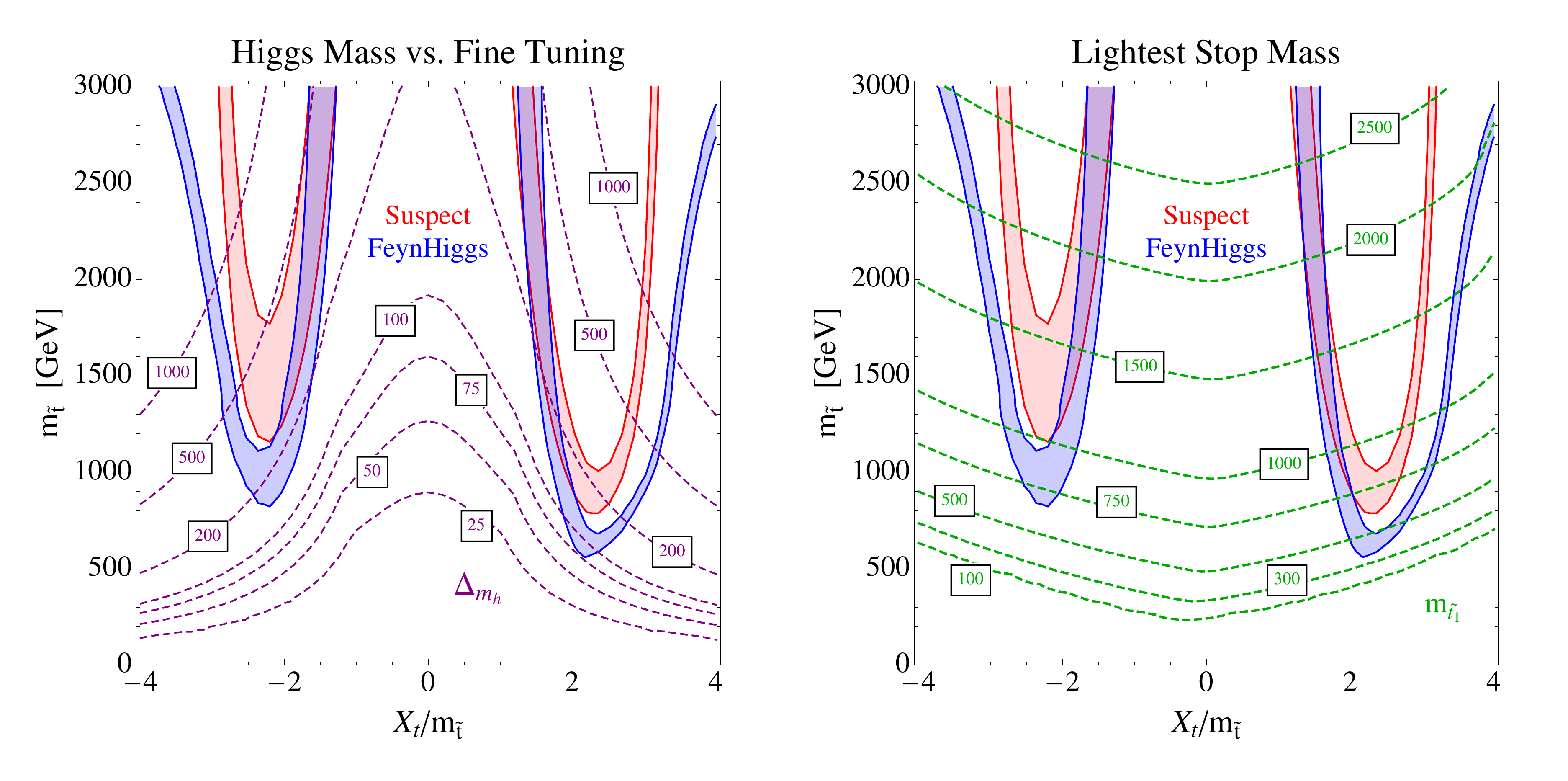} 
\caption{\label{fig:MSSM}
Contours of $m_h$ in the MSSM as a function of a common stop mass $m_{Q_3} = m_{u_3} = m_{\tilde t}$ and the stop mixing parameter $X_t$, for $\tan \beta = 20$.  The red/blue bands show the result from Suspect/FeynHiggs for $m_h$ in the range 124--126~GeV.  The left panel shows contours of the fine-tuning of the Higgs mass, $\Delta_{m_h}$, and we see that $\Delta_{m_h} > 75 (100)$ in order to achieve a Higgs mass of 124 (126) GeV.  The right panel shows contours of the lightest stop mass, which is always heavier than 300 (500) GeV when the Higgs mass is 124 (126) GeV.
}
\end{figure}

We use the Suspect~\cite{suspect} and FeynHiggs~\cite{feynhiggs} packages to calculate the Higgs mass, which include the full one-loop and leading two-loop contributions. In Figure \ref{fig:MSSM} we give the $m_h=124$ and 126~GeV contours in the $(X_t, m_{\tilde t})$ plane, with Suspect shown in red and FeynHiggs shown in blue.  For both curves, the axes are consistently defined in the $\overline{\text{DR}}$ renormalization scheme.  The left and right-handed top squark mass parameters are taken equal, $m_{Q_3} = m_{u_3}$, since the Higgs mass depends only mildly on the ratio. As we shall show, this choice results in the lowest fine-tuning for a given $m_{\tilde t}$, since the stop contribution to fine-tuning is dominated by the largest soft mass. The loop contribution depends slightly on the choice of some of the other SUSY parameters: we have fixed all gaugino masses to 1~TeV, the Higgsino mass to $\mu = 200$~GeV, and $m_A = 1$~TeV.  We find that the Suspect and FeynHiggs results have considerable differences.  The two programs use different renormalization prescriptions, 
and we take the difference between the two programs as a rough estimate of the theoretical uncertainty in the Higgs mass calculation.  For an earlier comparison, see~\cite{Allanach:2004rh}.  The uncertainty should be reduced if one takes into account the results of recent three-loop calculations~\cite{3loop}, although this is beyond the scope of our work.  For a detailed discussion of the two-loop calculations, see for example~\cite{Heinemeyer:2004ms}.  Fortunately, the two programs agree to within a factor of two on the necessary stop mass in the maximal mixing regime: $m_{\tilde t} =500-1000$~GeV for $X_t \sim \sqrt{6} m_{\tilde t}$ and $m_{\tilde t} \sim 800-1800$~GeV for $X_t \sim - \sqrt{6} m_{\tilde t}$, for a Higgs mass in the 124--126~GeV range.

We now consider the degree of fine-tuning~\cite{CERN-TH-4825/87,hep-ph/0007265,SusyFineTuning,hep-ph/0602096,GoldenMaxMix} necessary in the MSSM to accommodate a Higgs of 125 GeV.  We have just seen that rather heavy stops are necessary in order to boost the Higgs to 125 GeV using the loop correction.  The (well-known) problem is that heavy stops lead to large contributions to the quadratic term of the Higgs potential, $\delta m_{H_u}^2$,
\be \label{eq:1looptune}
\delta m_{H_u}^2 = - \frac{3 y_t^2}{8 \pi^2} \left( m_{Q_3}^2 + m_{u_3}^2 + | A_t |^2 \right)  \ln \left( \frac{\Lambda} {m_{\tilde t}} \right),
\ee
where $\Lambda$ is the messenger scale for supersymmetry breaking.
If $\delta m_{H_u}^2$ becomes too large the parameters of the theory must be tuned against each other to achieve the correct scale of electroweak symmetry breaking.  We see from equation~\ref{eq:1looptune} that large stop mixing also comes with a cost because $A_t$ induces fine-tuning.  At large $\tan \beta$, $X_t \approx A_t$, and maximal mixing ($|A_t|^2 = 6 m_{\tilde t}^2$) introduces the same amount of fine-tuning as doubling both stop masses in the unmixed case.

In order to quantify the fine-tuning~\cite{hep-ph/0602096}, it is helpful to consider a single Higgs field with a potential
\be \label{eq:toypotential}
V  = m_{H}^2 |h|^2 + \frac{\lambda_h}{4} |h|^4.
\ee
Extremizing the potential we see that the physical Higgs mass, $m_h$, is related to the quadratic term of the potential by $m_h^2 = \lambda_h v^2 = -2 m_H^2$.  The amount of fine-tuning is determined by the size of the Higgs mass relative to the size of corrections to the quadratic term of the potential.  In the MSSM at large $\tan \beta$, the Higgs vev is in the $H_u$ direction, $m_h$ corresponds to the Higgs mass, $m_H$ corresponds to $m_{H_u}$, and $\lambda_h$ is determined by the $D$-terms at tree-level and is logarithmically sensitive to the stop mass at one-loop.  We generalize to more than one Higgs field (2 in the MSSM and 3 in the NMSSM) by considering the sensitivity of the Higgs mass eigenvalue to variations of the fundamental parameters of the theory.  This is closely related to variations of the electroweak VEV, $v^2 = m_h^2/\lambda_h$, which is also often taken as a measure of fine-tuning.

\begin{figure}[t!]
\begin{center} \includegraphics[width=0.5\textwidth]{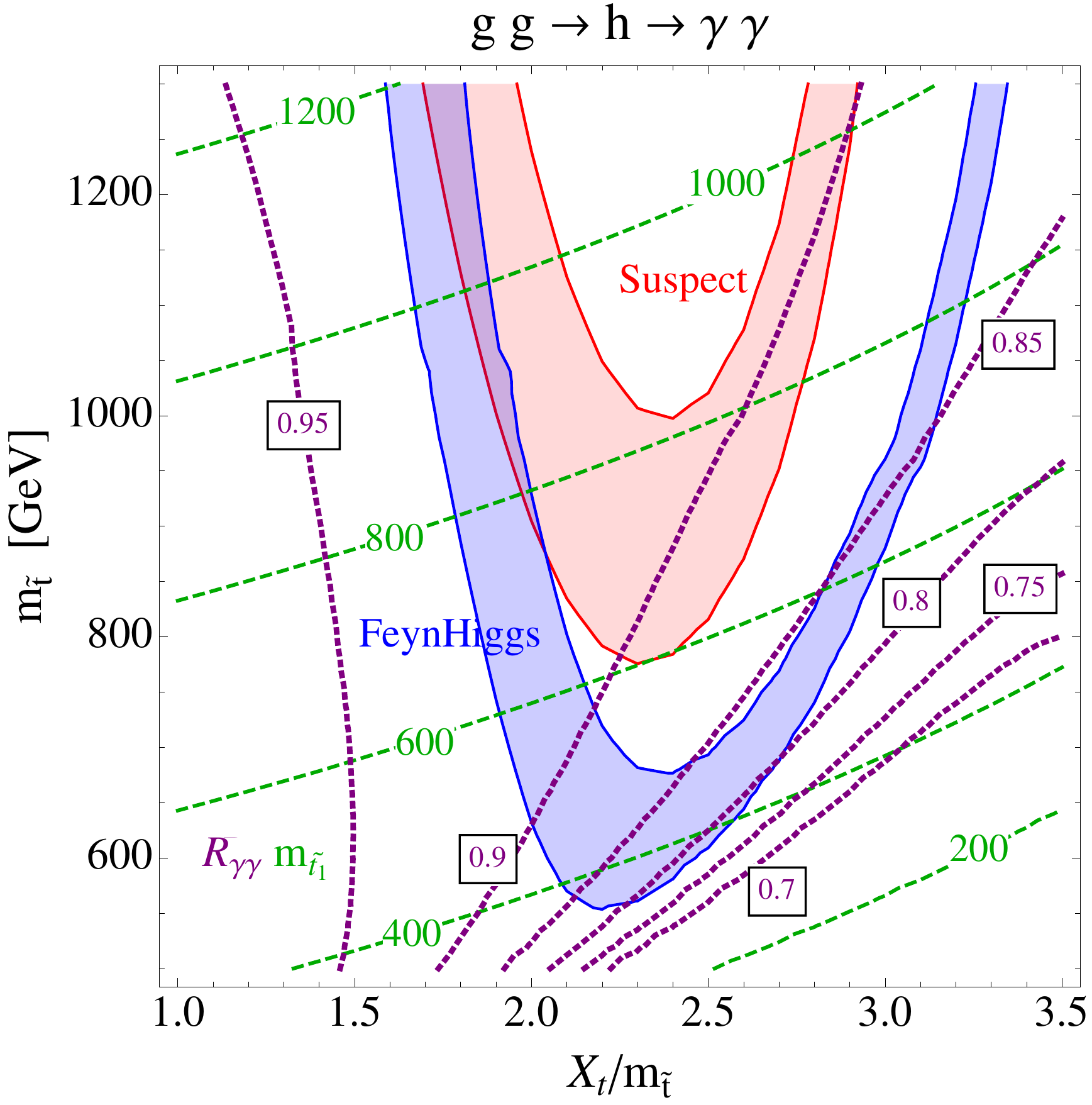} \end{center}
\caption{\label{fig:MSSMrate}
A blowup of the maximal mixing regime, $X_t \sim 2 m_{\tilde t}$, in the MSSM, with $\tan\beta = 20$ and $m_A = 1$ TeV.  The purple contours show $R_{\gamma \gamma}$, the ratio of $\sigma(g g \rightarrow h) \times \mathrm{Br}(h \rightarrow \gamma \gamma)$ in the MSSM to the Standard Model, computed with FeynHiggs.  The one-loop contribution from stops depletes the rate to be $\sim80 - 95\%$ of the SM rate. Had we chosen non-degenerate squark soft masses, this effect could be larger, at the cost of increased fine-tuning.
The other contours are the same as the right side of Figure~\ref{fig:MSSM}. 
}
\end{figure}

The dashed purple lines on the left panel of Figure~\ref{fig:MSSM} show contours of the fine-tuning parameter, $\Delta_{m_h}$, which we define to be the maximum logarithmic derivative of the Higgs boson mass with respect to the fundamental parameters, $p_i$,
\be \label{eq:finetune}
\Delta_{m_h} \, = \, \max_i\left|\frac{\partial \ln m_h^2}{\partial \ln p_i}\right| ,
\ee
where we take the fundamental parameters, defined at the messenger scale $\Lambda$, to be $\mu$, $B\mu$, $m_{Q_3}^2$, $m_{u_3}^2$, $A_t$, $m_{H_u}^2$, $m_{H_d}^2$.  We compute equation~\ref{eq:finetune} at tree-level and also include the one-loop leading log contribution to $m_{H_u}^2$, given by equation~\ref{eq:1looptune}, which allows us to relate the value of $m_{H_u}^2$ at the cutoff to its value at the weak scale.   For a 125 GeV Higgs mass the fine-tuning is smallest near maximal mixing, but even here the fine-tuning is severe, with $\Delta_{m_h} > 100 (200)$ for $X_t >0 (<0)$. Deviating away from maximal mixing, the squark masses quickly become multi-TeV in order to raise the Higgs mass to 125 GeV, and the fine-tuning is dramatically increased.  Furthermore, we stress that the fine-tuning has been computed for an extremely low value of $\Lambda=10$~TeV for the messenger scale.  For high-scale mediation schemes, such as gravity mediation, the fine-tuning is an order of magnitude worse.   The dashed green lines of the right panel of Figure~\ref{fig:MSSM} show contours of the lightest top squark mass, which can be as low as 400~GeV at maximal mixing but can rise to over a TeV with only a mild increase in fine-tuning.  

In Figure \ref{fig:MSSMrate} we show one of the regions of large stop mixing in the $(X_t, m_{\tilde t})$ plane with an expanded scale.  The red, blue and green contours are the same as in Figure \ref{fig:MSSM} and the dashed purple lines show contours of $R_{\gamma \gamma}$, the size of $\sigma(g g\rightarrow h)\times \mathrm{Br} (h \rightarrow \gamma \gamma)$ in the MSSM, computed using FeynHiggs and normalized to its value in the Standard Model. Here we have chosen $m_A = 1$ TeV, so that non-decoupling affects the rate at $< 3\%$. This rate is depleted relative to the SM because stop loops lower the Higgs coupling to gluons when the stops have a large mixing angle~\cite{Djouadi:2005gj}.  In this region the suppression of the $\gamma \gamma$ signal for a 125 GeV Higgs varies from about 0.8 to a little over 0.9.  The theoretical uncertainty on $\sigma(g g \rightarrow h)$ in the SM~\cite{arXiv:1101.0593} is about 10\%, and so a suppression at the lower end of this range may therefore be observable after enough statistics are accumulated.

\section{A Higgs Mass near 125 GeV in the NMSSM}
\label{sec:NMSSM}

We found above that a Higgs boson mass of 125~GeV introduces considerable fine-tuning into the MSSM.  This motivates us to go beyond the MSSM and look for a more natural theory of the Higgs sector.  A promising alternative is the Next-to-Minimal Supersymmetric Standard Model~\cite{Ellwanger:2009dp}, where a new singlet superfield couples to the Higgs in the superpotential, $\lambda S H_u H_d$.  The singlet coupling to the Higgs can contribute to the Higgs mass, potentially reducing the fine-tuning relative to the MSSM~\cite{hep-ph/0509244,NMSSMtuning,hep-ph/0612133, arXiv:0712.2903}.  In this section, we require the theory to remain perturbative up to the scale of gauge coupling unification, which requires $\lambda \lesssim 0.7$ at the weak scale. In the next section, we will consider the $\lambda$-SUSY scenario, where $\lambda$ is allowed to be larger.

We take the Higgs-sector of the superpotential to be,
\be \label{eq:nmssmW}
W \supset \lambda \, S H_u H_d + \hat \mu \, H_u H_d + \frac{M_S}{2} S^2.
\ee
We have included an explicit $\mu$-term and an explicit supersymmetric mass for $S$~\cite{MAD/PH/429}, in contrast to many studies of the NMSSM which focus on the scenario with no dimensionful terms in the superpotential.  We define the parameter $\mu = \hat \mu + \lambda \left< S \right>$, which acts as the effective $\mu$-term and sets the mass of the charged Higgsino.

We also include the following soft supersymmetry breaking terms,
\be \label{eq:nmssmV}
V_{\rm soft} \supset   m_{H_u}^2 |H_u|^2 + m_{H_d}^2 |H_d|^2 + m_S^2 |S|^2  + \left(B\mu \, H_u H_d  +  \lambda A_\lambda \, S H_u H_d   + \mathrm{h.c.}\right).
\ee
For simplicity, we have not included the trilinear interaction $S^3$ in the superpotential or scalar potential because we do not expect its presence to qualitatively change our results.  We neglect CP phases in this work and take all parameters in equations~\ref{eq:nmssmW} and~\ref{eq:nmssmV} to be real.

\begin{figure}[t!]
\includegraphics[width=1\textwidth]{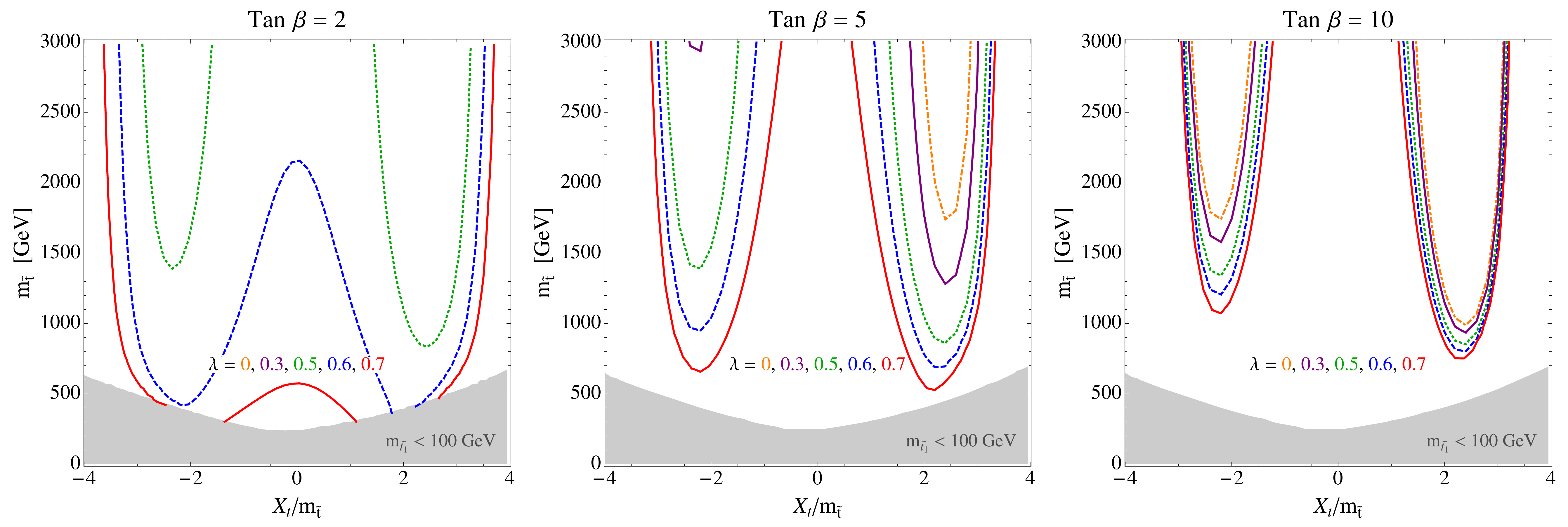} 
\caption{\label{fig:NMSSM2D}
Contours of $m_h = 125$~GeV in the NMSSM, taking $m_{Q_3} = m_{u_3} = m_{\tilde t}$ and varying $\tan \beta = 2, 5, 10$ from left to right, and varying $\lambda$ within each plot.  We add the tree-level Higgs mass (with NMSSM parameters chosen to maximize it) to the two-loop stop contribution from Suspect.  The tree-level Higgs mass is largest at lower values of $\tan \beta$ and larger values of $\lambda$, where only modestly heavy stops, $m_{\tilde t} \sim 300$~GeV, are needed to raise the Higgs to 125~GeV.
Heavy stops are still required for lower values of $\lambda$ and larger values of $\tan \beta$.
} 
\end{figure}

In this section, we focus on the scenario where the lightest CP-even scalar is mostly doublet, with doublet-singlet mixing not too large.  The lightest CP-even scalar mass that results from the above potential is bounded from above at tree-level~\cite{Ellwanger:2009dp},
\be \label{eq:massbound}
({m_h}^2)_{\rm tree} \le m_Z^2 \cos^2 2 \beta + \lambda^2 v^2 \sin^2 2 \beta.
\ee
Since we take the lightest scalar to be dominantly doublet, this is a bound on the Higgs mass.\footnote{It is also interesting to consider the case where the lightest eigenstate is dominantly singlet.  Then, singlet-doublet mixing can increase the mass of the dominantly doublet eigenstate~\cite{arXiv:0712.2903}.}  The first term is the upper bound in the MSSM, while the second term is the contribution from the interaction involving the singlet.
The above bound is saturated when the singlet is integrated out with a large supersymmetry breaking mass, $m_S^2 > M_S^2$~\cite{hep-ph/0509244}, which, in practice, can be realized with $m_S$ several hundreds of GeV\@.  For large enough values of $\lambda$, the second term dominates the tree-level mass. The $\lambda$ term grows at small $\tan \beta$, and this means that the largest Higgs mass is achieved with low $\tan \beta$  and as large $\lambda$ as possible.  Plugging in $\lambda = 0.7$, we find that $({m_h}^2)_{\rm tree}$ is always smaller than 122~GeV.  

\begin{figure}[t!]
\includegraphics[width=1\textwidth]{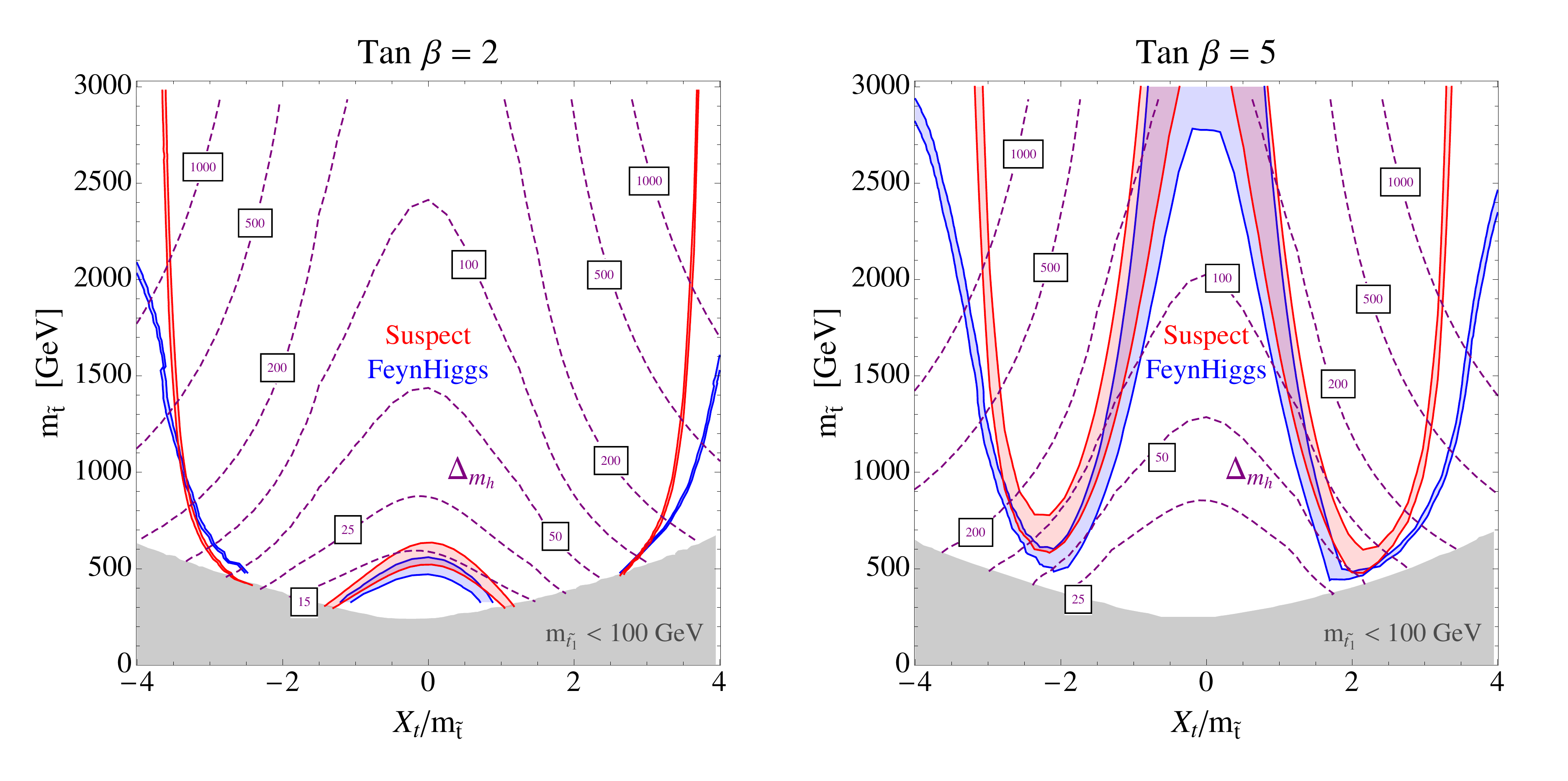} 
\caption{\label{fig:NMSSMtune}
Contours of Higgs mass fine-tuning, $\Delta_{m_h}$, in the NMSSM with the maximal value of $\lambda = 0.7$ for $\tan \beta = 2$ and 5, moving from left to right, with $m_{Q_3} = m_{u_3} = m_{\tilde t}$ and $m_A = 500$~GeV.  Contours of $m_h = 124$ and 126~GeV are overlaid, including loop corrections from Suspect and FeynHiggs.  When $\tan \beta = 2$ the tuning can be low,  $\Delta_{m_h} \lesssim 15$, while for $\tan \beta = 5$ heavier stop masses are required because the tree-level Higgs mass is lower.
}
\end{figure}

Because the tree-level contribution is insufficient to raise the Higgs mass to 125~GeV, we also consider the loop corrections to the Higgs mass arising from stops.  In Figure~\ref{fig:NMSSM2D}, we show contours of $m_h=125$~GeV, in the stop mass/mixing plane, with $\tan \beta =2,5,10$ and varying $\lambda$ between 0 and 0.7.  We take the tree-level mass to saturate the bound of equation~\ref{eq:massbound} and we add to it the one and two loop contribution from stops using Suspect, taking degenerate stop soft masses, $m_{Q_3} = m_{u_3}$.  Here, and for the rest of this section, we have set $\mu = 200$~GeV and we fix $B\mu$ by taking the MSSM-like pseudoscalar mass to be 500 GeV, in the limit of no mixing with the singlet-like pseudoscalar.  Suspect includes only the MSSM contribution, and this means that we are neglecting the one-loop contribution proportional to $\lambda^2$, which is a reasonable approximation since $\lambda < y_t$.  For low $\tan \beta$ and $\lambda$ close to 0.7, the lightest stop becomes tachyonic near maximal mixing.  Furthermore, for sub-maximal stop mixing, the stops are light enough to give $\mathcal{O}(1)$ corrections to $\sigma(g g \rightarrow h)$; however, these corrections may take either sign, depending on the size of the mixing~\cite{Djouadi:2005gj}, which is relatively unconstrained by naturalness for large $\lambda$. Furthermore, the stop mass quickly rises as $\lambda$ is decreased or as $\tan \beta$ is increased, decoupling this effect.  

Next we consider the fine-tuning in the NMSSM when $m_h = 125$~GeV.  As in section~\ref{sec:MSSM}, we measure fine-tuning in terms of the sensitivity of the Higgs mass to the parameters of the theory using the maximum logarithmic derivative of equation~\ref{eq:finetune}.  We consider derivatives with respect to the MSSM parameters $\mu$, $B\mu$, $m_{Q_3}^2$, $m_{u_3}^2$, $A_t$, $m_{H_u}^2$, and $m_{H_d}^2$, defined at the cutoff, which we conservatively take to be the low scale of $\Lambda = 10$~TeV.
We also take derivatives with respect to the NMSSM parameters $m_S^2$, $M_S$, and $A_\lambda$.  We include the one-loop contribution of $m_S^2$ to $m_{H_{u,d}}^2$, which is proportional to $\lambda^2$.  The result is shown in the stop mass/mixing plane in Figure~\ref{fig:NMSSMtune}, taking $\lambda$ as large as possible, 0.7, and taking $\tan \beta=2$ on the left and $\tan \beta = 5$ on the right.   For $\tan \beta = 2$, where the tree-level Higgs mass is larger, we find that the fine-tuning is typically around 1/15 for moderate to low stop mixing.  The tuning is mild because the stops are light, and in fact for low stop mass the tuning is dominated by the choice of $\mu$, and can even be lowered to the 10\% level if $\mu$ is lowered to 100 GeV.  For $\tan \beta=5$, the tree-level Higgs mass is smaller and heavier stops are required to raise the Higgs to 125~GeV.  This results in more fine-tuning, and for $\tan \beta = 5$, we find that maximal mixing is required to avoid multi-TeV stops, and the fine-tuning is always worse than $\sim$2--3\%.  As $\tan \beta$ rises, the fine-tuning as a function of stop masses and mixing quickly reverts to that of the MSSM.

\begin{figure}[h!]
\includegraphics[width=1\textwidth]{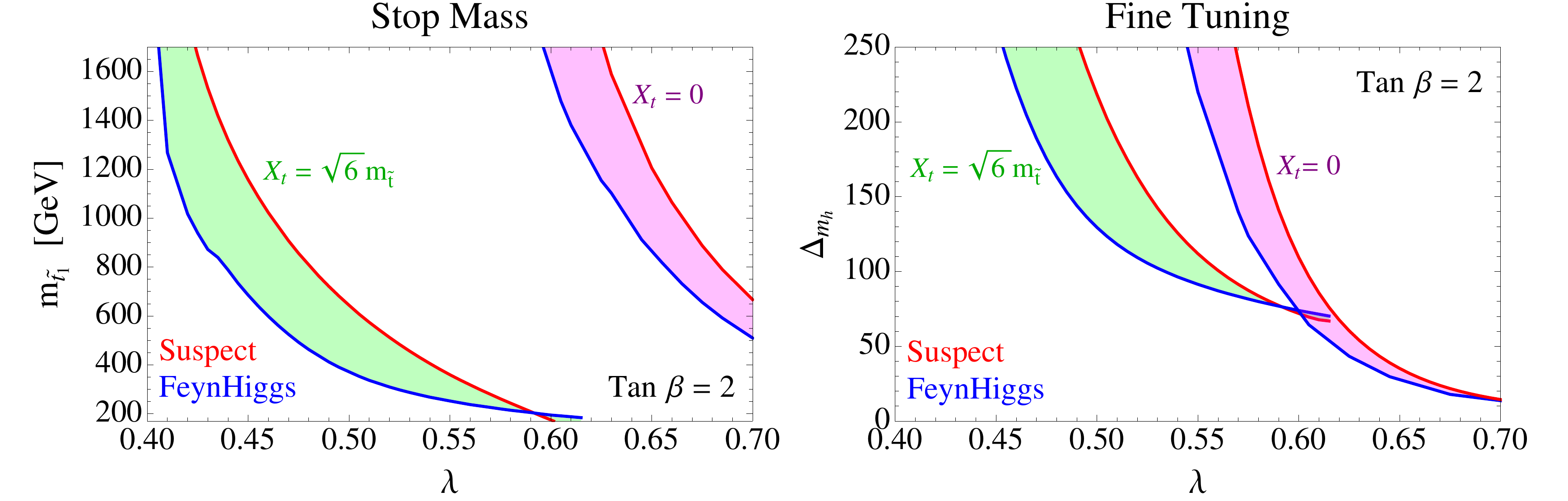} 
\caption{\label{fig:NMSSMlambda}
The necessary stop mass (left) and fine-tuning (right) in order to achieve a Higgs mass of 125~GeV in the NMSSM, as a function of $\lambda$. We see that larger values of $\lambda$ allow for lighter stops and much less fine-tuning.  We consider two cases for the stop mixing: (1) maximally mixed stops and (2) zero mixing.  We cut off the plot for maximally mixed stops when $m_{\tilde{t}_1} \sim m_t$.  For both plots, the loop corrections are computed using Suspect and FeynHiggs, and we fix $\tan \beta = 2$.
}
\end{figure}

Finally, we conclude this section by looking at how the necessary stop mass and fine-tuning depends on $\lambda$, which is shown in figure~\ref{fig:NMSSMlambda}.  We fix the Higgs mass to 125~GeV and $\tan \beta = 2$, and we look at the stop mass given separately by FeynHiggs and Suspect for maximal stop mixing and no stop mixing, with degenerate stop soft masses.    We see that the necessary stop mass drops dramatically from multi-TeV at $\lambda < 0.5$ to hundreds of GeV near $\lambda$ of 0.7.  We also show the fine-tuning in the Higgs mass as a function of $\lambda$ and see that the fine-tuning drops from 1/1000 at low $\lambda$ to 1/100 near $\lambda = 0.5-0.6$ for maximal mixing,  and finally to close to 1/15 near $\lambda = 0.7$. Note that, while the required stop mass for a given Higgs mass can be dramatically smaller at maximal mixing, the fine-tuning is significantly worse than in the case of no mixing, since $A_t$ contributes to the running of $m_{H_u}$. Clearly, the most interesting regime of the NMSSM is where $\lambda$ is as large as possible.  This motivates us, in the next section, to relax the requirement that the theory remain perturbative until the scale of gauge coupling unification, and to consider the implications of a 125~GeV Higgs for theories with $\lambda > 0.7$.

\section{A Higgs Mass near 125 GeV in $\lambda$-SUSY}
\label{sec:lambdasusy}

In the previous section we found that, at low $\tan\beta$, increasing $\lambda$ in the NMSSM improves naturalness and allows for much lighter stops. Motivated by this, we now consider values of $\lambda~>~0.7$, larger than is allowed for perturbative unification.  However, in certain UV completions such as Fat Higgs models~\cite{fathiggs}, successful gauge coupling unification can occur even with $\lambda > 0.7$ and a non-perturbative sector well below unified scales. In this section we study values of $\lambda$ up to 2, beyond which the running value of $\lambda^2(10 \, \rm{TeV})$ becomes of order $4 \pi$, and non-perturbative effects are likely to upset precision electroweak data~\cite{hep-ph/0607332}.  Many features of this $\lambda$-SUSY framework have been studied~\cite{hep-ph/0607332, LambdaSusyPheno, LambdaSusyLightScalar}, but always with the SM-like Higgs boson heavier than about 160 GeV.

Here we study the theory defined by the interactions of Eqs.~\ref{eq:nmssmW} and~\ref{eq:nmssmV} from the previous section, but with large $\lambda$.  We begin by considering the Higgs mass, which is naively of order $\lambda v \sim$ 200--300 GeV in the limit of small $\tan \beta$ (see Eq.~\ref{eq:massbound}).  However, this estimate neglects mixing between the Higgs and the CP even singlet within $S$, which would be a good approximation in the limit of large singlet soft mass, $m_S \rightarrow \infty$.  However, this limit cannot be taken consistently with naturalness, because the singlet scalar soft mass affects the $H_u$ and $H_d$ soft masses through the one-loop RGEs,
\be \label{eq:mHudRGEs}
\frac{d m_{H_d}^2}{dt} &=& \lambda^2 \frac{m_S^2}{8\pi^2} + \cdot\cdot\cdot~, \\
\frac{d m_{H_u}^2}{dt} &=& \lambda^2 \frac{m_S^2}{8\pi^2} + \frac{3}{8\pi^2} y_t^2 \left(m_{Q_3}^2 + m_{u_3}^2 + |A_t|^2\right) + \cdot\cdot\cdot~.
\ee
Naturalness requires the singlet scalar to be relatively light, $m_S \lesssim 1$~TeV, and so singlet-doublet mixing between the CP even mass eigenstates must be considered, as in the mass matrix of Eq.~\ref{eq:h-smassmixing} from the introduction.  The off-diagonal mixing terms arise from the soft $A$-term and the cross-terms in the $F$-terms of the potential, 
\be \label{eq:Lmix}
\mathcal{L} \supset \lambda v \left[(A_{\lambda} + M)\cos\beta - 2\mu\sin\beta\right] s \, h_u + \lambda v \left[(A_{\lambda} + M)\sin\beta - 2\mu\cos\beta\right]s \, h_d,
\ee
and they become large as $\lambda$ is raised and $m_S$ is dropped. Level-splitting then drives the smallest mass eigenstate lighter, as shown in Figure~\ref{fig:LambdaSUSYHiggs}, allowing for much smaller Higgs masses than would be expected from the singlet-decoupling limit.

\begin{figure}[h!]
 \begin{center} \includegraphics[width=0.5\textwidth]{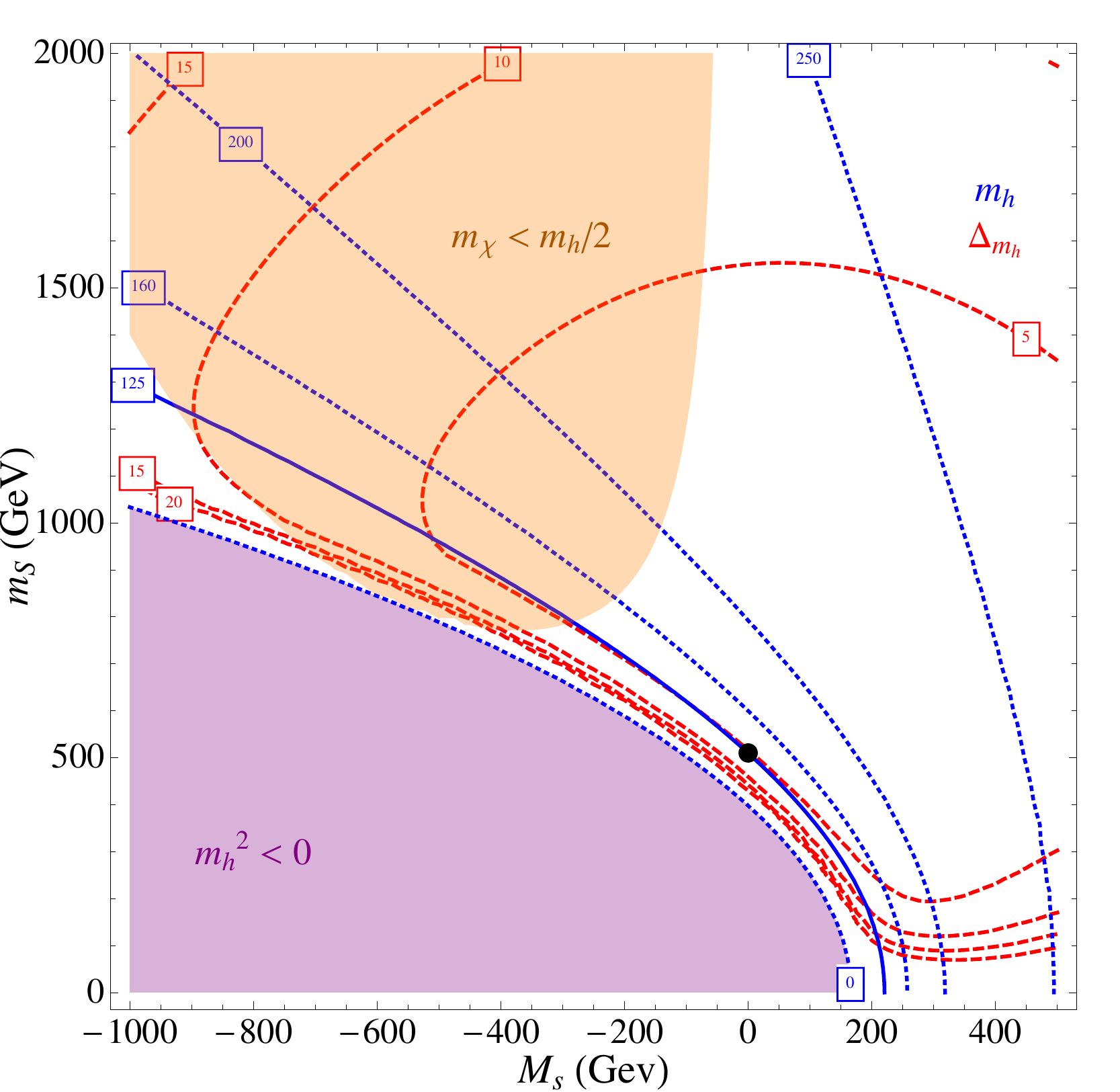} \end{center}
\caption{\label{fig:MsmSTuning}
The Higgs mass in $\lambda$-SUSY varying the singlet supersymmetric mass, $M_S$, and soft mass, $m_S$.  The Higgs mass contours are shown in blue, contours of Higgs fine-tuning, $\Delta_{m_h}$, are shown in red, and the region where the Higgs is tachyonic, due to Higgs-singlet mixing, is shown in purple.  The fine-tuning is increased when the Higgs mass drops, however, a Higgs mass of 125~GeV is achieved in a region of low fine-tuning,  $\Delta_{m_h} \sim 5$.  The orange region is where the lightest neutralino is lighter than half the Higgs mass, and in this region the Higgs would dominantly decay invisibly.
}
\end{figure}

\begin{figure}[h!]
 \begin{center} \includegraphics[width=1\textwidth]{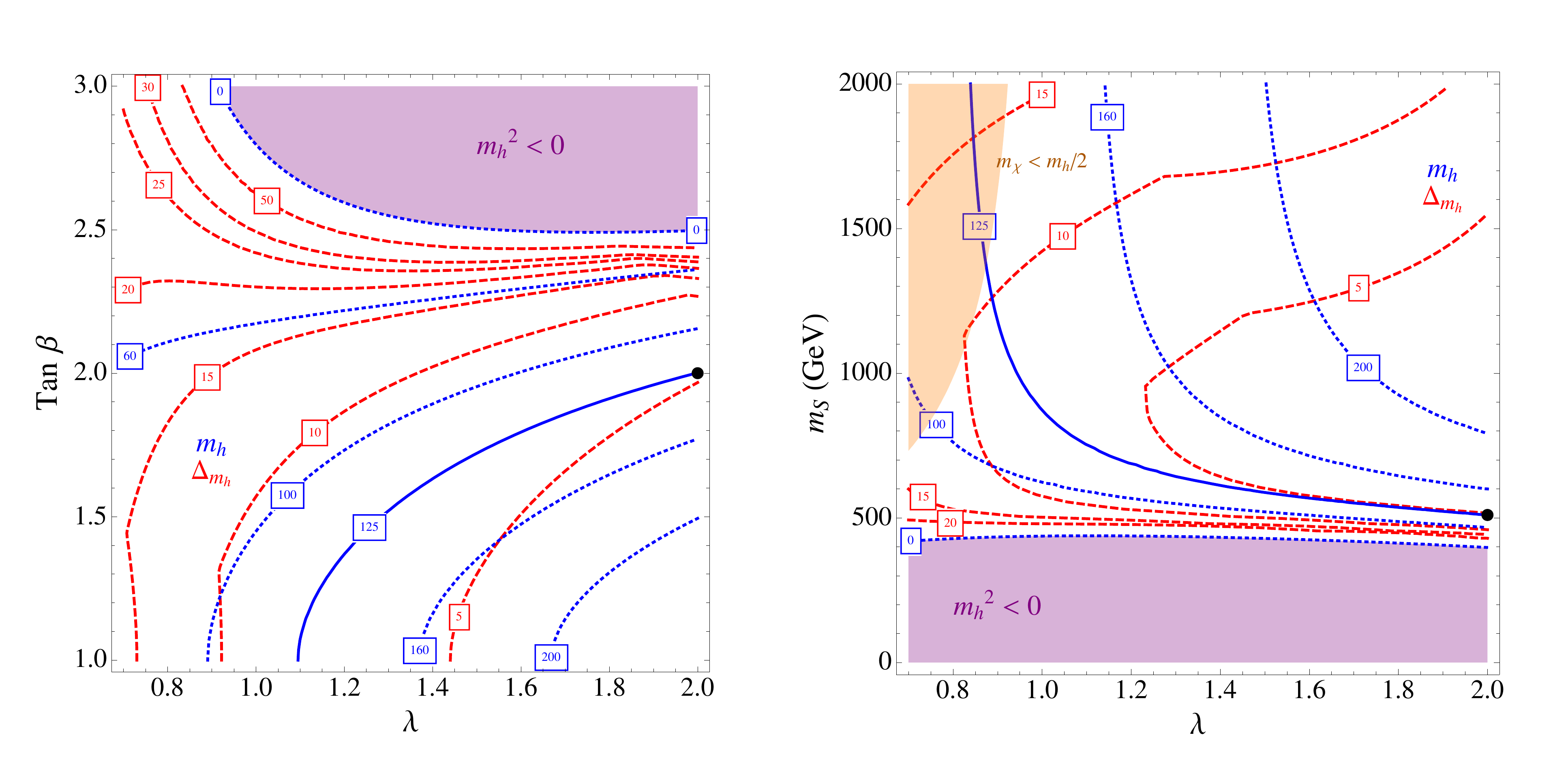} \end{center}
\caption{\label{fig:lambdaTuning}
The Higgs mass and fine-tuning contours, $\Delta_{m_h}$ in $\lambda$-SUSY.  On the left, we vary $\lambda$ and $\tan \beta$ and on the right we vary $\lambda$ and the singlet soft mass, $m_S$.  The rest of the parameters are fixed as in table~\ref{tab:bench}.   We find that there is a preference for large $\lambda$, small $\tan \beta$, and moderate values of the singlet soft mass, $m_S \sim 500$~GeV.  Overall, there is a large region of parameter space where a Higgs mass of 125~GeV is consistent with very mild tuning, $\Delta_{m_h} \sim 5$.    Within the purple region, the Higgs is driven tachyonic due to Higgs-singlet mixing, and in the orange region on the right plot, there is a light neutralino and the Higgs dominantly decays invisibly.  
}
\end{figure}

We now proceed to calculate the Higgs mass at tree-level, which is a good approximation when $\lambda \sim 2$ because $\lambda^2 v^2$ is large relative to one-loop corrections from stops.  For the purpose of illustration we choose a reference point in parameter space, shown in Table~\ref{tab:bench}, with a maximal value of $\lambda = 2$ and other parameters shown in the table.  At this point, we see that $m_S \sim 500$~GeV leads to a Higgs mass of 125~GeV due to the level-splitting.  In Figure~\ref{fig:MsmSTuning} we show contours of the Higgs mass in the $(m_S,M_S)$ plane and in Figure~\ref{fig:lambdaTuning} the contours of the Higgs mass in the $(\lambda, \tan \beta)$ and $(\lambda, m_S)$ planes.   For each figure, we hold the parameters not being varied fixed to the values of the benchmark point in Table~\ref{tab:bench}. Several Higgs mass contours are shown in blue, with the 125~GeV contour solid and the others dotted.   The black dot shows the reference point.  
In the right panel of Figure~\ref{fig:MsmSTuning} and in Figure~\ref{fig:lambdaTuning}, as the singlet scalar mass $m_S$ is reduced below the reference point, the Higgs mass rapidly drops to zero and the tachyonic purple region with $m_h^2 < 0$ is reached, demonstrating that the Higgs mass is being affected by singlet-doublet Higgs mixing, as illustrated in Figure~\ref{fig:LambdaSUSYHiggs}.

Obtaining a Higgs mass much smaller than $\lambda v$ requires a tuning in the level splitting. Thus, a light Higgs could signal an additional fine-tuning not captured by standard measures which seek to quantify cancellations inherent in the weak scale. In order to capture both effects, we once again consider logarithmic derivatives of the Higgs mass, rather than the weak scale, as a measure of fine-tuning. Here we include one-loop LL contributions from the stop masses and $\lambda$. Since $\lambda$ increases rapidly with energy, we improve the LL contribution by integrating its RGE up to a messenger scale of $\Lambda = 10$ TeV, obtaining 
\be \label{eq:lambdaRGimproved}
\delta m_{H_{d}}^2(\Lambda) &\simeq& \frac{m_S^2}{4}\ln\left(1 - \frac{\lambda^2}{2\pi^2}\ln\frac{\Lambda}{\sqrt{m_{Q_3} m_{u_3}}}\right), \\
\delta m_{H_{u}}^2(\Lambda) &\simeq& \frac{m_S^2}{4}\ln\left(1 - \frac{\lambda^2}{2\pi^2}\ln\frac{\Lambda}{\sqrt{m_{Q_3} m_{u_3}}}\right) \nonumber \\
&& \qquad \qquad \qquad -\frac{3 y_t^2}{8\pi^2} \left(m_{Q_3}^2 + m_{u_3}^2 + |A_t|^2\right) \ln\frac{\Lambda}{\sqrt{m_{Q_3} m_{u_3}}}. 
\ee

\begin{table}[h!]
\begin{center}
\begin{tabular}{|c|c||cc|}
\hline
\multicolumn{2}{|c||}{parameters} & \multicolumn{2}{c|}{properties} \\
\hline \hline
$\lambda =2$ & $\tan \beta = 2$ & $m_h = 125\unit{GeV}$ & $\theta_{hs}= 0.12$	\\
$\mu = 200\unit{GeV}$&$M_S=0\unit{GeV}$ &  \multicolumn{2}{c|}{$m_{h_{2,3}} = 521, 662\unit{GeV}$} \\
$m_S = 510\unit{GeV} $&  $m_{H^+} =470\unit{GeV}$&\multicolumn{2}{c|}{ $m_{A_{1,2}} = 579,617\unit{GeV}$}\\
\cline{1-2}
\multicolumn{2}{|c||}{$m_{Q_3} = m_{u_3}=500\unit{GeV}$} & \multicolumn{2}{c|}{$ \Delta_{m_h}=5.2$} \\
\multicolumn{2}{|c||}{$A_t, A_\lambda=0$} &\multicolumn{2}{c|}{$ \xi_{b \bar b, t \bar t, \gamma \gamma, WW} = (0.27,1.03,0.79,0.84) $}\\ 
\multicolumn{2}{|c||}{}&\multicolumn{2}{c|}{$R_{\gamma \gamma} = 1.67$ \quad $R_{WW}=1.79$  \quad $R_{bb}=0.46$}\\
\hline
\end{tabular}
\end{center}
\caption{\label{tab:bench}
A benchmark point in $\lambda$-SUSY with a large $\lambda$ of 2 and a 125~GeV Higgs boson mass, which results from Higgs-singlet mixing.  The parameters are shown to the left and various masses, mixing angles, and phenomenologically relevant Higgs couplings are shown to the right.  The Higgs boson mass is not fine-tuned relative to the fundamental parameters, $\Delta_{m_h} \sim 5$.  Here, the $R_i$ parameters represent the ratio of $\sigma \times \mathrm{Br}$, relative to the SM, with $\sigma$ corresponding to gluon fusion for the $\gamma \gamma$ and $WW$ final states and associated $Z/W+h$ production for $h \rightarrow bb$.
}
\end{table}

Contours of the fine-tuning parameter $\Delta_{m_h}$ are shown as red dashed lines in Figures~\ref{fig:MsmSTuning} and~\ref{fig:lambdaTuning}. In each plot, the region with $m_h$ approaching zero has a high density of red contours, and this area is highly fine-tuned, demonstrating the sensitivity of the Higgs mass to the parameters entering its mixing angle with the singlet. However, a large range of relatively light Higgs masses lies just outside this area, as expected from the mild slope in Figure~\ref{fig:LambdaSUSYHiggs}. In fact, the reference point lies near the edge of a large region of the $(\lambda, \tan \beta, m_S, M_S)$ parameter space where $\Delta_{m_h} \sim 5$.  The $\lambda$-SUSY theory has a large region with $m_h = 125$~GeV that is less fine-tuned than the NMSSM. In a portion of this region, which we have shaded in orange in Figure~\ref{fig:MsmSTuning} and the right panel of Figure~\ref{fig:lambdaTuning}, the lightest neutralino mass is less than one half of the Higgs mass, so that Higgs decays to neutralinos becomes kinematically accessible.  Due to the large coupling, one expects that such invisible decays will occur with an order one branching ratio as soon as they are allowed~\cite{futurework}.

\begin{figure}[h!]
 \begin{center} \includegraphics[width=0.5\textwidth]{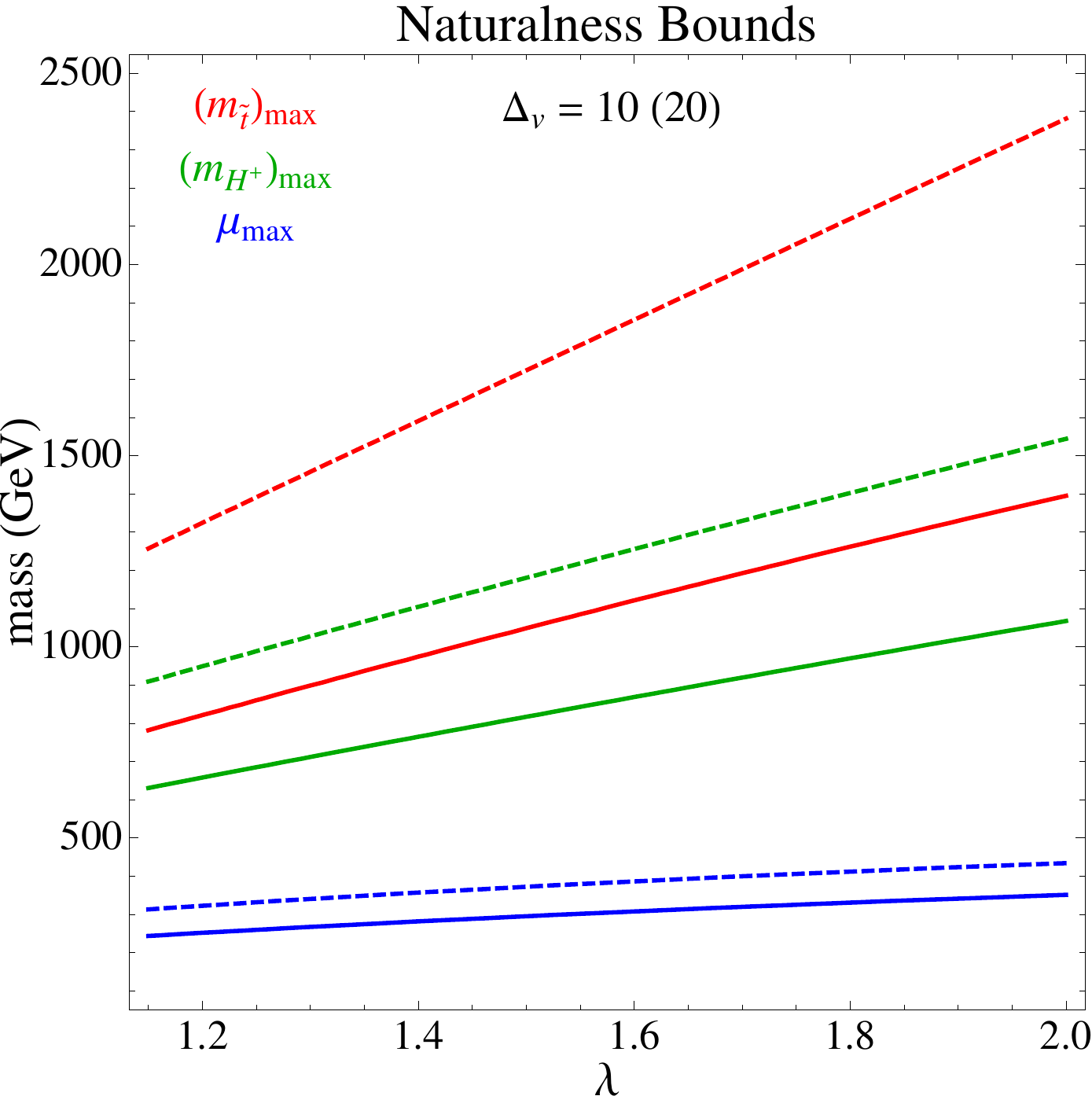} \end{center}
\caption{\label{fig:maxmass}
The maximum values of the stop, charged Higgs, and Higgsino masses, before the fine-tuning of the electroweak vev becomes worse than 10\% (5\%) is shown with solid (dashed) lines.  We vary $\lambda$ along the horizontal axis and for each choice of $\lambda$ (and $\mu$, $m_{H^+}$), we choose $M_S$ in such a way as to fix the Higgs mass to 125~GeV.  We see that larger values of $\lambda$ allow for heavier sparticles while maintaining naturalness.  For $\lambda=2$, the stops can be as heavy as 1.4~TeV, the charged Higgs can be 1~TeV, and the Higgsinos can be around 350~GeV.
}
\end{figure}

It is clear from Figure~\ref{fig:lambdaTuning} that the fine-tuning is only a mild function of $\lambda$. However, large $\lambda$ has a very important effect: it protects the Higgs mass from heavy sparticle corrections, decreasing the Higgs mass sensitivity to the sparticle spectrum. At large $\lambda$, the fine-tuning is dominated by the sensitivity to the parameters entering the singlet-doublet mixing, while at small $\lambda$ (and large $m_S$) the mixing becomes less important, and the fine-tuning comes from stop loops correcting $m_{H_u}^2$. For a given fine-tuning the stop mass must be lower for a lower value of~$\lambda$.

In Figure~\ref{fig:maxmass} we show the maximum stop, chargino, or charged Higgs mass required to raise $\Delta_{v}$ to 10 and 20, fixing the light Higgs mass to be 125~GeV. Here we have chosen to show $\Delta_{v}$ because it results in more conservative mass values than $\Delta_{m_h}$. Raising $\lambda$ to 2 allows the stop mass to be twice as large for the same fine-tuning as would a $\lambda$ of 1. Recall that the MSSM and NMSSM with a stop mass at 1.4 TeV are at least an order of magnitude more fine-tuned. We also see from this plot that, as in the MSSM, the chargino and the charged Higgs masses should not be too large. We may understand this by considering the minimization conditions for the potential in the limit of small singlet-doublet mixing, as was done in~\cite{hep-ph/0509244}. One finds
\be \label{eq:Vmin}
\lambda^2 v^2 &=& \frac{2 B_\mu}{\sin 2\beta} - (2|\mu|^2 + m_{H_u}^2 + m_{H_d}^2) \\
&=& \frac{2 B_\mu}{\sin 2\beta} - m_{H^+}^2 + m_W^2,
\ee
so that neither $\mu$ nor $m_{H^+}$ should be far above $\lambda v$.

\begin{figure}[h!]
 \begin{center} \includegraphics[width=1\textwidth]{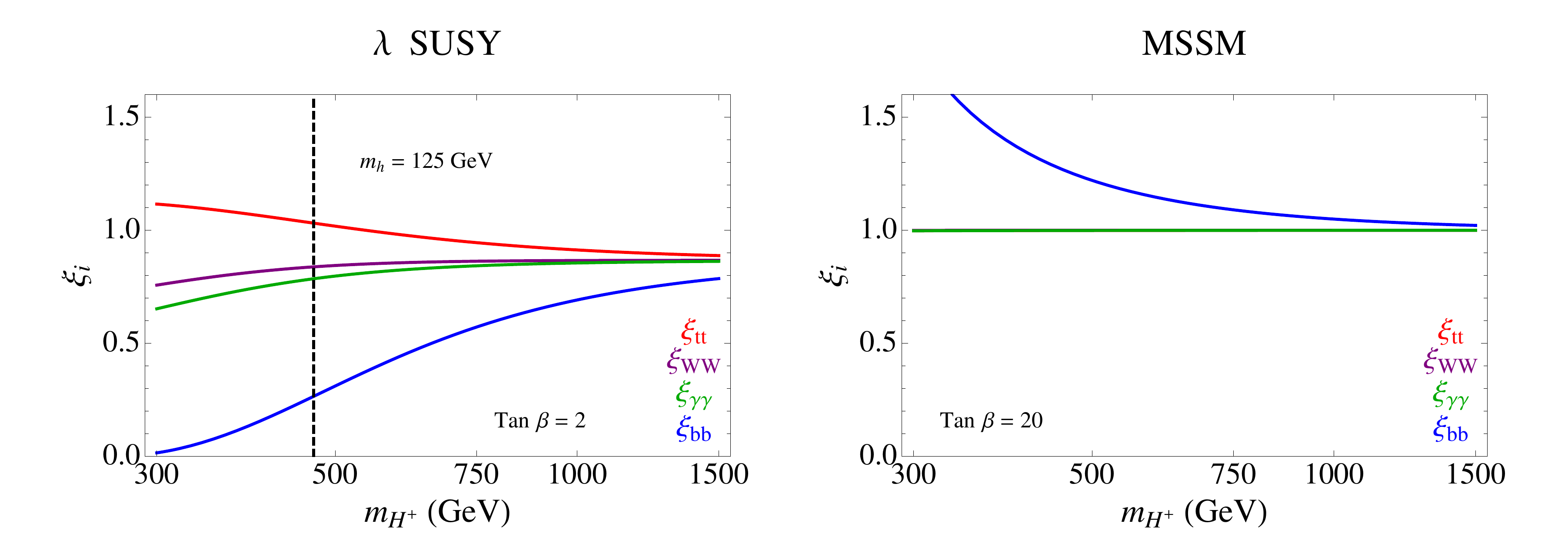} \end{center}
\caption{\label{fig:couplings}
The ratio of Higgs couplings squared relative to the Standard Model for $b \bar b$, $t \bar t$, $W^- W^+$ and $\gamma \gamma$ as a function of the charged Higgs mass, $m_{H^+}$.  $\lambda$-SUSY is shown to the left and the MSSM is shown to the right.  In $\lambda$-SUSY the couplings are computed at tree-level, and the Higgs mass is a function of $m_{H^+}$; in the MSSM we approximate the one-loop correction to the couplings~\cite{Djouadi:2005gj}, given that the stop contribution raises the Higgs mass to 125~GeV.  We see that in $\lambda$-SUSY, unlike the MSSM, the Higgs coupling to the bottom quark drops dramatically away from the decoupling limit, leading to a depleted Higgs width and an enhanced $\gamma \gamma$ signal.  The $\lambda$-SUSY parameters, other than the charged Higgs mass, are as in table~\ref{tab:bench}; for the MSSM we choose $\tan \beta = 20$.
}
\end{figure}

Keeping the charged Higgs relatively light has interesting phenomenological consequences in $\lambda$-SUSY: the non-decoupling effects turn off slower than in the MSSM, resulting in modified Higgs couplings and branching ratios. Expanding in powers of $v/(m_A,m_{H^+})$ the light Higgs coupling to $b\bar{b}$ in the MSSM normalized to the SM is\footnote{The formula given here includes the approximate one-loop contribution to the Higgs mixing angle from the stops, with the assumption that they only affect the $m_{H_u}^2$ element of the Higgs mass matrix~\cite{Djouadi:2005gj}.}
\be \label{eq:MSSMyb}
\xi_{bb} \equiv \frac{y_b^2}{(y_b^2)_{\text{SM}}} &\approx& 1 + 2\left(1 - \left(\frac{m_Z}{m_h}\right)^2\cos2\beta\right) \left(\frac{m_h}{m_A}\right)^2 \\
&\rightarrow& 1 + |\sin4\beta|\tan\beta \left(\frac{m_Z}{m_A}\right)^2 \text{ at tree level},
\ee
while in $\lambda$-SUSY we have
\be \label{eq:LSUSYyb}
\xi_{bb} &\approx& 1 - |\sin 4\beta| \tan \beta \left(\frac{\lambda v}{m_H^+}\right)^2,
\ee
neglecting corrections from singlet-doublet mixing. In $\lambda$-SUSY the non-decoupling effect is a factor of $\sim 2\lambda^2/g^2$ larger and takes the opposite sign as compared to the MSSM, tending to reduce the Higgs coupling to $b\bar{b}$. We show the ratio of light Higgs couplings to various particles in the MSSM and in $\lambda$-SUSY relative to those in the SM in Figure~\ref{fig:couplings} as a function of $m_{H^+}$. The MSSM Higgs coupling to $b\bar{b}$ can be enhanced by an order one amount as the charged Higgs mass approaches the $b\rightarrow s\gamma$ limit near $\sim 300$ GeV~\cite{Misiak:2006zs}, while the couplings to $WW$, $\gamma\gamma$, and $t\bar{t}$ remain nearly unperturbed by decoupling effects. In contrast, it can be seen that the $b\bar{b}$ coupling may be decreased dramatically in $\lambda$-SUSY, reaching a value of 0.3 relative to the SM at our benchmark point from Table~\ref{tab:bench}. The depletion of the coupling to $WW$ is not as severe because it first appears at order $(\lambda v/ m_{H^+})^4$ in the expansion. Furthermore, the $\xi_i$ do not asymptote to 1 because of the singlet-doublet mixing, which tends to deplete all couplings uniformly. 

\begin{figure}[h!]
 \begin{center} \includegraphics[width=1\textwidth]{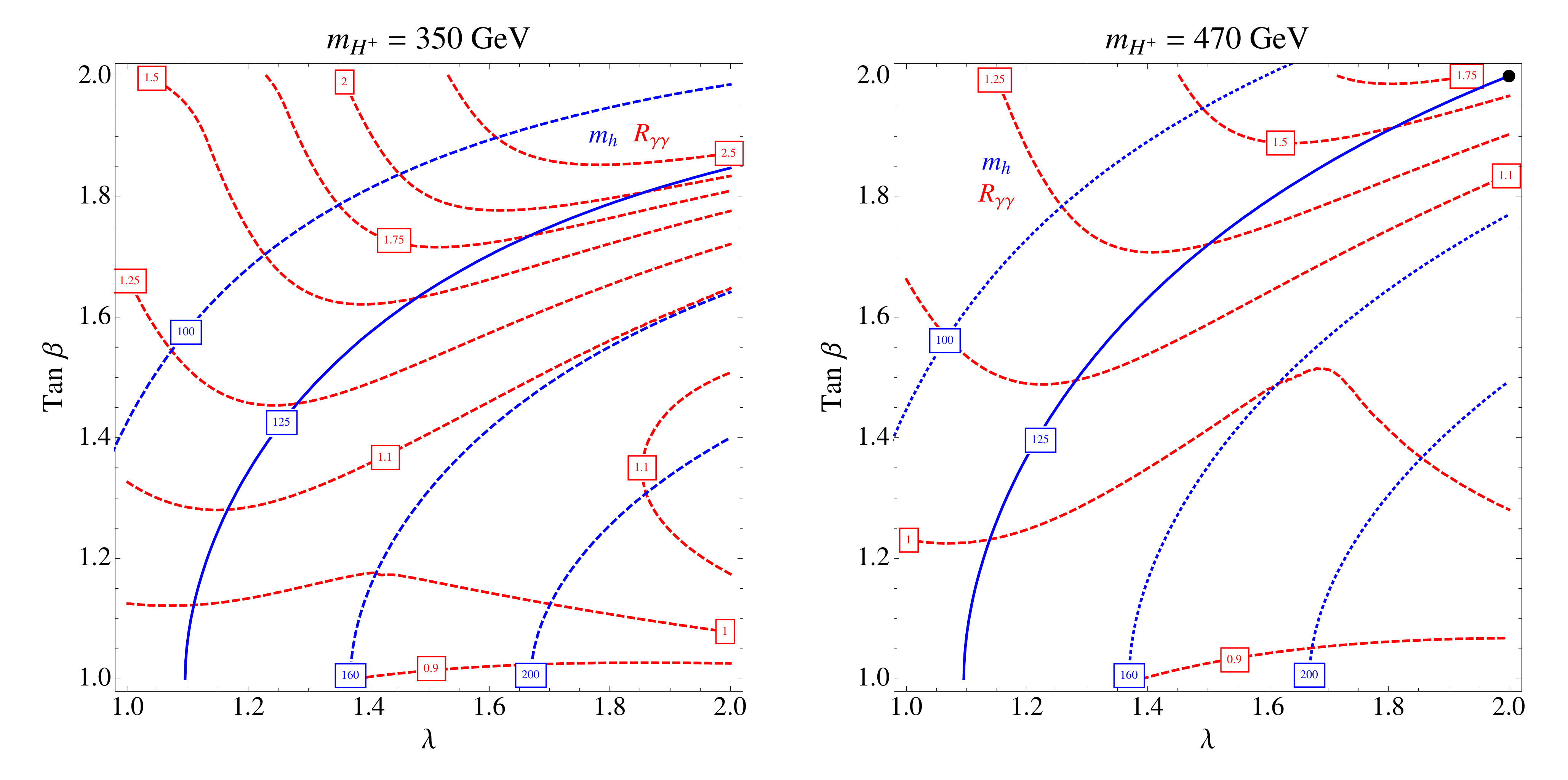} \end{center}
\caption{\label{fig:sigmaBr}
The ratio $R_{\gamma \gamma}$ of $\sigma \times Br$ in $\lambda$-SUSY relative to the SM for the process $g g \rightarrow h \rightarrow \gamma \gamma$.  The red contours show $R_{\gamma \gamma}$ and the blue contours show the Higgs mass in the $\lambda, \tan \beta$ plane.  We see that this process generically has a larger rate in $\lambda$-SUSY than in the SM, this is due to the depletion of the Higgs coupling to bottom quarks in the non-decoupling limit.  The left and right panels correspond to charged Higgs masses of 350 and 470~GeV, respectively.
}
\end{figure}

For a SM-like Higgs at 125~GeV, decays to $b\bar{b}$ contribute $58\%$~\cite{hep-ph/9704448} to the full width. Thus, a depletion of the $b\bar{b}$ coupling can generate a large increase in the branching ratios to other final states relative to a SM or MSSM Higgs. In Figure~\ref{fig:sigmaBr} we show contours of $\sigma\times$Br($gg\rightarrow h\rightarrow\gamma\gamma$) relative to the SM in the ($\lambda,\tan\beta$) plane for two values of $m_{H^+}$.  We compute the modified branching ratios by weighting the partial widths of the SM Higgs~\cite{hep-ph/9704448} by the $\xi_i$ factors discussed above. As expected from Eq~\ref{eq:LSUSYyb}, the enhancement to $R_{\gamma\gamma}$ grows with $\lambda$, and can be greater than 1.5 in a large region of parameter space with low fine-tuning and a light Higgs. The enhancement turns off quickly as $m_{H^+}$ is raised, but $m_{H^+}$ cannot become too large without inducing fine-tuning, so that naturalness prefers a larger-than-SM rate to $\gamma\gamma$.  Since doublet-doublet mixing has a comparable effect on $\xi_{WW}$ as to $\xi_{\gamma\gamma}$, the depleted Higgs width should similarly enhance the rate to $WW^*$, $R_{WW} \sim R_{\gamma\gamma}$.  Large enhancements to the $\gamma\gamma$ rate can also be found in the NMSSM, although typically for Higgs masses lighter than 125~GeV~\cite{NMSSMgammagamma}.  Even in the MSSM, stop loops can have a similar effect on the Higgs production cross section and therefore the overall rate to photons~\cite{Djouadi:2005gj}; thus, in order to distinguish these non-decoupling effects from stop contributions, the rate to $b\bar{b}$ and $\tau\tau$ must be measured as well, since they should be depleted relative to the MSSM by the decrease in the Higgs coupling. At the benchmark point described in Table~\ref{tab:bench}, we find $R_{bb} = R_{\tau\tau} = .46$ when the Higgs is produced either by vector boson fusion or associated $Z/W + h$ production, which are the relevant channels for Higgs discovery in the $b\bar{b}$ and $\tau\tau$ modes. It is also possible to distinguish these effects by measuring the vector boson fusion rate to photons, which will also be enhanced by non-decoupling effects, whereas stop loops mainly affect the Higgs coupling to gluons.

\begin{figure}[h!]
 \begin{center} \includegraphics[width=0.65\textwidth]{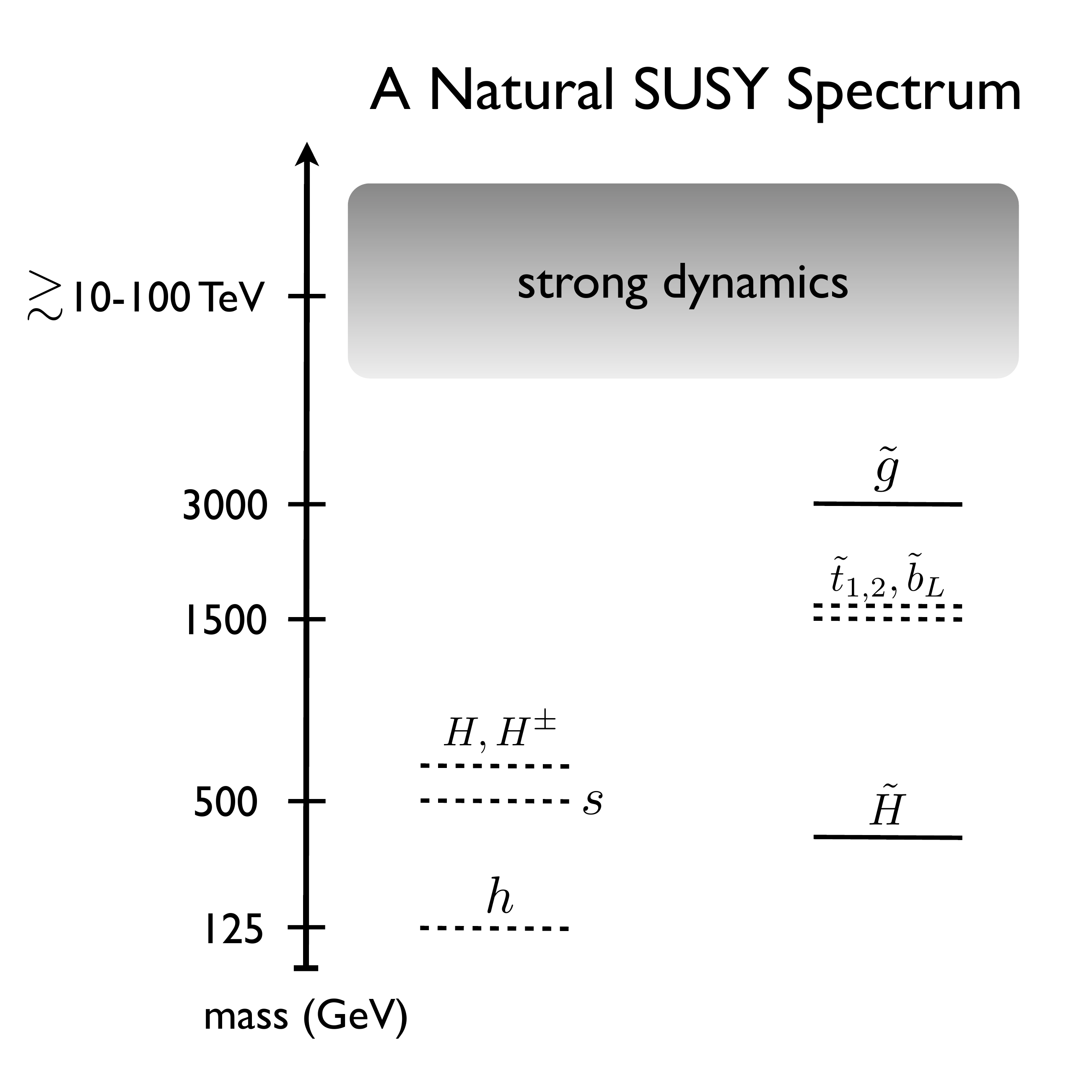} \end{center}
\caption{\label{fig:NaturalScheme}
An example of a natural SUSY spectrum in $\lambda$SUSY with $\lambda \sim 2$.  The fine-tuning of the Higgs mass, and electroweak symmetry breaking, can remain milder than 10\% with the Higgsinos at 350 GeV, the stops at 1.5 TeV, and the gluino at 3 TeV.  Mixing between the Higgs and the singlet lowers the Higgs mass to 125~GeV.
}
\end{figure}

\section{Conclusions}
\label{sec:concl}

The LHC experimental collaborations have presented data that can be interpreted as the first evidence for a Higgs boson with a mass of about 125~GeV~\cite{AtlasTalk,CMSTalk}.  In this paper, we have studied the implications of a 125~GeV Higgs boson for naturalness in supersymmetric theories.  We considered three scenarios: the MSSM, the NMSSM with a coupling $\lambda$ that can remain perturbative until the scale of gauge coupling unification, and $\lambda$-SUSY, where $\lambda$ can be larger.  Our main results concerning naturalness are,

\begin{itemize}

\item In the MSSM, maximal mixing is required to avoid multi-TeV stops.  Fine-tuning is at the 1\% level or worse  with a low mediation scale of 10~TeV, and an order of magnitude more with a high mediation scale.

\item The NMSSM can accommodate a 125~GeV Higgs with only $\sim5-10\%$ tuning if the mediation scale is low and stop mixing is non-maximal.  In order to achieve such mild fine-tuning, the NMSSM is pushed to the edge of its parameter space, with low $\tan \beta \lesssim 2$ and large $\lambda \sim 0.7$, so that $\lambda$ is very nearly non-perturbative at the GUT scale. 

\item $\lambda$-SUSY presents a highly natural theory of a 125~GeV Higgs, with tuning in the range of 10-20\% for a large portion of its parameter space.  The Higgs mass can be 125~GeV in theories with large $\lambda$, because it is naturally driven light by Higgs-singlet mixing.  Alternatively, a 125~GeV Higgs mass can be achieved in theories with somewhat smaller $\lambda$ or larger $\tan \beta$, if Higgs-singlet mixing is not an important effect.

\end{itemize}

We have discussed a number of  phenomenological consequences of the $\lambda$-SUSY theory with large $\lambda \sim 2$.  Even though the Higgs mass can be as low as 125~GeV, the stops can be very heavy, about 1.5~TeV, before they introduce fine-tuning into electroweak symmetry breaking.  In Figure~\ref{fig:NaturalScheme} we give an example of such a natural superparticle spectrum.  This possibility presents a new twist on the null supersymmetry results: maybe superparticles are above the 7 TeV reach of the LHC because the Higgs potential is protected by a large value for $\lambda$. Of course, since the tree-level contributions are large in $\lambda$-SUSY, the stops are not required to be heavy in order to raise the Higgs mass. Thus it is also possible that the superparticle spectrum is about to be discovered. We have also found that $\lambda$-SUSY has the possibility of interesting non-decoupling effects.  Mixing between the two doublets depletes the coupling of the lightest Higgs to bottom quarks (the opposite of how non-decoupling usually works in the MSSM), enhancing the $\gamma \gamma$ and $WW$ rates and depleting the branching ratios to $b$'s and $\tau$'s.  In $\lambda$-SUSY, non-SM Higgs branching ratios may present the first experimental clue for supersymmetry, instead of the direct discovery of sparticles. 

Of course, the experimental results remain at a preliminary stage and whether or not the Higgs boson is really present at 125~GeV will be fleshed out by data presented in the coming year.  We conclude by discussing how our results might be modified if the Higgs signal at 125~GeV goes away and the Higgs is instead discovered with a lower mass in the window between the LEP limit of 114~GeV~\cite{Barate:2003sz} and the current LHC limit of about 130~GeV.  A lower Higgs mass would have crucial implications for the MSSM and NMSSM.  Recall that the necessary stop mass, and therefore the degree of fine-tuning, depends exponentially on the Higgs mass.  If the Higgs is found closer to the LEP limit of 114~GeV, the MSSM, and a larger portion of the NMSSM parameter space, would look a lot more appealing than it does if the Higgs has a mass of 125~GeV.   On the other hand, our results pertaining to $\lambda$-SUSY carry over, basically unmodified, for any Higgs mass within the currently allowed window.  A Higgs mass of 125~GeV is most naturally achieved at tree-level in theories with large $\lambda$, and the amount of fine-tuning is not terribly sensitive to whether the mass is close to 114~GeV or 125~GeV.  New data over the coming year will determine if the current excess is robust and thus whether naturalness points towards non-minimal supersymmetry.

\section*{Acknowledgments}

We thank Nima Arkani-Hamed, Asimina Arvanitaki, Yasunori Nomura, Matthew Reece, Pietro Slavich, and Giovanni Villadoro for helpful conversations.
This work was supported in part by the Director, Office of Science, Office 
of High Energy and Nuclear Physics, of the US Department of Energy under 
Contract DE-AC02-05CH11231 and by the National Science Foundation under 
grant PHY-1002399.  J.T.R. is supported by a fellowship from the Miller Institute for Basic Research in Science.

\end{document}